\documentclass[fleqn]{aa}
\usepackage{mathtools}
\usepackage{txfonts}
\usepackage{natbib}
\usepackage{graphicx}
\usepackage{epstopdf}
\usepackage[switch]{lineno}
\usepackage{pgf}
\usepackage{soul}
\definecolor{lightgreen}{HTML}{B7F774}
\sethlcolor{yellow}

\newcommand{\ds}{\displaystyle}

\begin{document}

\title{Brightness temperature constraints from interferometric visibilities}
\titlerunning{Brightness temperature from interferometric visibilities}

\author{Andrei Lobanov\inst{1,2}
}

    \institute{Max-Planck-Institut f\"ur Radioastronomie,
              Auf dem H\"ugel 69, 53121 Bonn, Germany
    \and
    Institut f\"ur Experimentalphysik, Universit\"at Hamburg, 
    Luruper Chaussee 149, 22761 Hamburg, Germany
  }
\authorrunning{Lobanov}

  \date{Received 1 October 2014 / Accepted 4 December 2014 }
 
  \abstract
  {The brightness temperature is an effective parameter that
describes the
    physical properties of emitting material in astrophysical
    objects. It is commonly determined by imaging and modeling the
    structure of the emitting region and estimating its flux density
    and angular size.}
  {Reliable approaches for visibility-based estimates of
    brightness temperature are needed for interferometric experiments
    in which poor coverage of spatial frequencies prevents successful
    imaging of the source structure, for example, in interferometric
    measurements made at millimeter wavelengths or with orbiting
    antennas.}
  {Such approaches can be developed by analyzing the
    relations between brightness temperature and visibility amplitude
    and its r.m.s. error.}
  {A method is introduced for directly calculating the lower and upper limits of
the    brightness temperature from visibility measurements.  The
    visibility-based brightness temperature estimates are shown to agree well with the image-based estimates obtained in the
    2\,cm MOJAVE survey and the 3\,mm CMVA survey, with good
    agreement achieved for interferometric measurements at spatial
    frequencies exceeding $\approx 2\times 10^8$.}
  {The method provides an essential tool for constraining brightness
    temperature in all interferometric experiments with poor imaging
    capability.}


   \maketitle
%

\section{Introduction}

Interferometric measurements offer a powerful tool for probing the finest
structures of emitting objects by extending the effective instrumental
diameter to the maximum distance (baseline length) between
individual elements of an interferometer. However, for interferometric
measurements made at extreme baseline lengths, imaging
the structure of the target object becomes increasingly limited owing to
incomplete sampling of the Fourier plane. This is often the case in
radio interferometric measurements made with very long baseline
interferometry (VLBI) at millimeter wavelengths \citep[{\em
  cf}.][]{doeleman+2012} and with space-ground interferometers such
as VSOP \citep{horiuchi+2004} or RadioAstron
\citep{kardashev+2013}. In these situations, more basic measurements of
flux density, $S_\nu$, and emitting area, $\Omega$, of the structure
can still be obtained ({\em i.e.}, from model fitting of the visibility
distribution) and can be combined to yield a brightness
temperature estimate. The latter can then be used as a generic indicator of the
physical conditions of the emitting material \citep[{\em
  cf}.][]{lobanov+2000,kovalev+2005,homan+2006,lee+2008}.

For the black-body spectrum in the Rayleigh-Jeans limit ($h\,\nu \ll
k\,T$), the brightness, $I_\nu$, is approximated by $I_\nu = 2 \nu^2
k\,T/c^2$, and the respective brightness temperature is $T_\mathrm{b}
= I_\nu c^2/(2\,k\,\nu^2)$, where $h$ are $k$ are the Planck and
Boltzmann constants, respectively, and $c$ is the speed of light.  In
terms of the measured quantities, $S_\nu$ and $\Omega$, the resulting
brightness is $I_\nu = S_\nu / \Omega =
S_\nu/[2\pi\,(1-\cos\rho_{d})]$, if the emitting region is a
uniformly bright circle of angular radius $\rho_\mathrm{d}$. For
small $\rho_\mathrm{d}$, the term $1- \cos\rho_\mathrm{d}$ is
approximated by $\rho_\mathrm{d}^2/2$, which yields $I_\nu \approx
S_\nu/(\pi\,\rho_\mathrm{d}^2)$. If the emitting region is unresolved,
$\rho_\mathrm{d}$ can be constrained by the resolution limit,
$\theta_\mathrm{lim}$, of the measurement, providing lower limits on
the brightness, $I_\nu \ge 4 S_\nu /(\pi\, \theta_\mathrm{lim}^2)$, and
brightness temperature, $T_\mathrm{b} \ge 2 S_\nu c^2 /(\pi\, k\, \nu^2
\theta_\mathrm{lim}^2)$.

In absence of information about the actual brightness distribution of
emission in a compact, marginally resolved region, it is often assumed
that it can be represented satisfactorily by a two-dimensional
Gaussian distribution described by a flux density $S_\mathrm{g}$ and
respective major and minor axes $\theta_\mathrm{maj}$ and
$\theta_\mathrm{min}$. This translates into $I_\nu = (4\,\ln\,2/\pi)\,
S_\mathrm{g}/(\theta_\mathrm{maj}\, \theta_\mathrm{min})$ and
$T_\mathrm{b} = [2\,\ln\,2/(\pi\,k)]\,S_\mathrm{g}\, c^2/(\nu^2\,
\theta_\mathrm{maj}\, \theta_\mathrm{min})$. These expressions are used
for the bulk of brightness temperature estimates based on
decomposition of the observed structure into one or more
two-dimensional Gaussian features (Gaussian components).  Several
other analytical patterns of brightness distribution patterns have
been employed to analyze different astrophysical objects
\citep[{\em cf}.][]{berger2003} such as resolved stars \citep{dyck+1998,ohnaka+2013}, young supernovae
\citep{marcaide+2009}, recurrent novae \citep{chesneau+2007},
protoplanetary disks \citep{malbet+2005}, or active galaxies
\citep{weigelt+2012}. In all of these cases, successful fitting of a
given brightness distribution pattern to visibility data is a
strong prerequisite for recovering structural and physical information
about the target object.

However, the most extreme cases of interferometric observations, such as millimeter and space VLBI measurements, often do not provide
enough data to warrant reliable model fitting owing to lack of short-baseline measurements and the complexity of the fine structure in most of the
targets. These observations require a different approach for estimating
the brightness temperature. In this paper, such an approach is
proposed based on individual visibility measurements and their
errors (which can be reliably estimated in most of the
measurements). The methodology of this approach is described in
Sect.~\ref{sc:method} and is tested with the visibility data from two
VLBA\footnote{Very Long Baseline Array of National Radio Astronomy
  Observatory, Socorro NM, USA; http://www.nrao.edu} observations of
the prominent compact radio sources 3C\,345 and NGC\,1052. Applications of
the visibility-based $T_\mathrm{b}$ estimates are discussed in
Sect.~\ref{sc:discussion} and are compared with the results from the 3\,mm
CMVA\footnote{Coordinated Millimeter VLBI Array, currently succeeded
  by the Global Millimeter VLBI Array;
  http://www3.mpifr-bonn.mpg.de/div/vlbi/globalmm/} and 2\,cm
MOJAVE\footnote{http://www.phyiscs.purdue.edu/astro/MOJAVE/} surveys
and in connection with an analysis of space VLBI and millimeter VLBI
surveys of compact radio sources. Additional potential applications of
the method to other types of astrophysical targets are also discussed,
and expressions for brightness temperature limits for several specific
brightness distribution patterns are presented in the appendix.

\section{Brightness temperature limits from interferometric visibilities}
\label{sc:method}

We consider an emitting region with a brightness distribution,
$I_\mathrm{r}$, observed instantaneously at a wavelength, $\lambda$,
by an interferometer consisting of two receiving elements (telescopes)
separated by a baseline distance, $B$. This observation corresponds to
measuring the Fourier transform of $I_\mathrm{r}$ at a single spatial
(Fourier) frequency $q = B/\lambda$ (also called {\em uv} distance or
{\em uv} radius).  It yields an interferometric visibility, $V =
V_\mathrm{q} e^{-i\,\phi_\mathrm{q}}$, described by its amplitude
$V_\mathrm{q}$ and phase $\phi_\mathrm{q}$, and their respective
errors $\sigma_\mathrm{q}$ and $\sigma_\phi$. Generally speaking,
$V_\mathrm{q}$ depends on the shape and angular extent of the
brightness distribution, and $\phi_\mathrm{q} = \phi_\mathrm{p} +
\phi_\mathrm{o}$ is a function of its position and geometry. The
position-dependent term of the phase, $\phi_\mathrm{p}$, is relative
and can always be zeroed by an appropriate shift applied to the
visibility (which is analogous to re-pointing the interferometer). The
geometry-dependent term of the phase, $\phi_\mathrm{o}$, depends on
the structure of the brightness distribution and its orientation with
respect to the projection of the interferometric baseline on the
picture plane. For a circularly symmetric or axially symmetric
brightness distribution, $\phi_\mathrm{o} \equiv 0$, independently of
the baseline orientation.

\subsection{Minimum brightness temperature}

Without {\em a priori} information about the specific structural shape of
$I_\mathrm{r}$, the symmetry assumption can be employed and the
angular extent of the emission can be estimated from $V_\mathrm{q}$
alone. This assumption is routinely used for size and brightness
temperature estimates made from interferometric data \citep[{\em
  cf}\,][]{lobanov+2000,kovalev+2005,lee+2008}.

Such estimates require knowledge of the {\em
  zero-spacing} visibility, $V_0$, and rely upon assumption of a specific
symmetric template for $I_\mathrm{r}$. For instance, for a circular
Gaussian distribution, the respective expression for $V_\mathrm{q}$ is
\[
V_\mathrm{q} = V_0 \exp\left(-\frac{\pi^2 \theta_\mathrm{r}^2 q^2}{4 \ln 2}\right)
\]
and it can be used for obtaining an estimate of the size,
$\theta_\mathrm{r}$, of the emitting region:
\begin{equation}
\theta_\mathrm{r} = \frac{2 \sqrt{\ln 2}}{\pi}\frac{\lambda}{B}\sqrt{\ln(V_0/V_\mathrm{q})}
\label{eq:thetar}
.\end{equation}
With this expression, the brightness temperature can be estimated from
\begin{equation}
T_\mathrm{b} = \frac{\pi}{2 k} \frac{B^2\, V_0}{\ln(V_0/V_\mathrm{q})}\,.
\label{eq:tbV0}
\end{equation}
Appendix A lists the respective expressions derived for several other patterns of brightness distributions commonly used in the analysis of astronomical data.

\begin{figure}[t!]
\begin{center}
\includegraphics[width=0.45\textwidth,bb=0 0 550 385,clip=true]{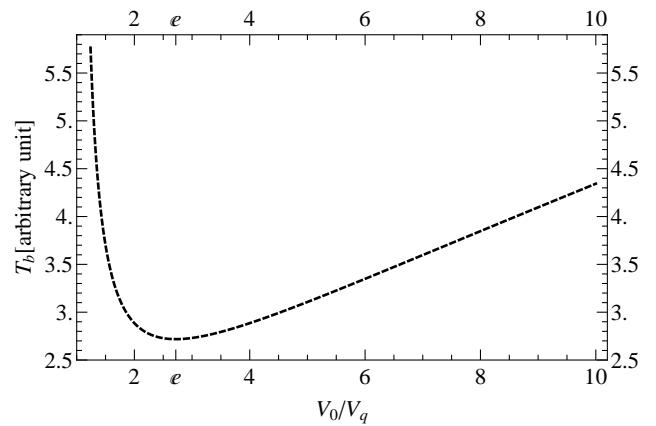}
\end{center}
\caption{Brightness temperature in relative units as a function of the
  ratio $V_0/V_\mathrm{q}$ between the zero-spacing flux density and
  visibility flux density measured at a given spatial frequency $q$,
  assuming that the brightness distribution is Gaussian. The lowest  value of the brightness temperature is realized with the ratio
  $V_0/V_\mathrm{q}=e$.}
\label{fg:VoVu}
\end{figure}

\begin{figure}[t!]
\includegraphics[height=0.37\textwidth]{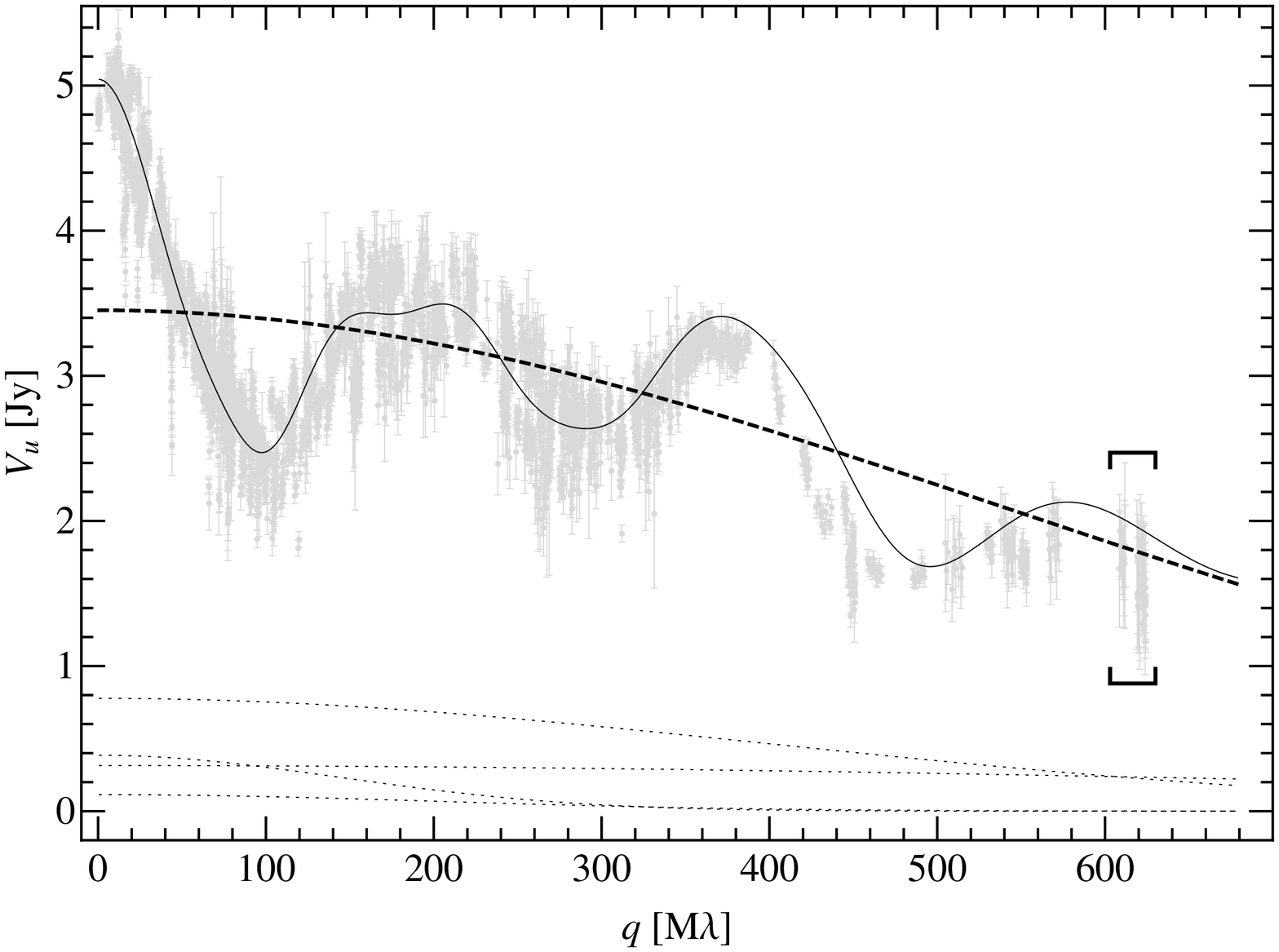}
\vskip-0.36\textwidth
\hbox to 0.465\textwidth{
\hfil
\includegraphics[height=0.15\textwidth,angle=-90,bb=67 166 510 629,clip=true]{fig02b.eps}
}
\vskip 0.19\textwidth
\includegraphics[height=0.37\textwidth]{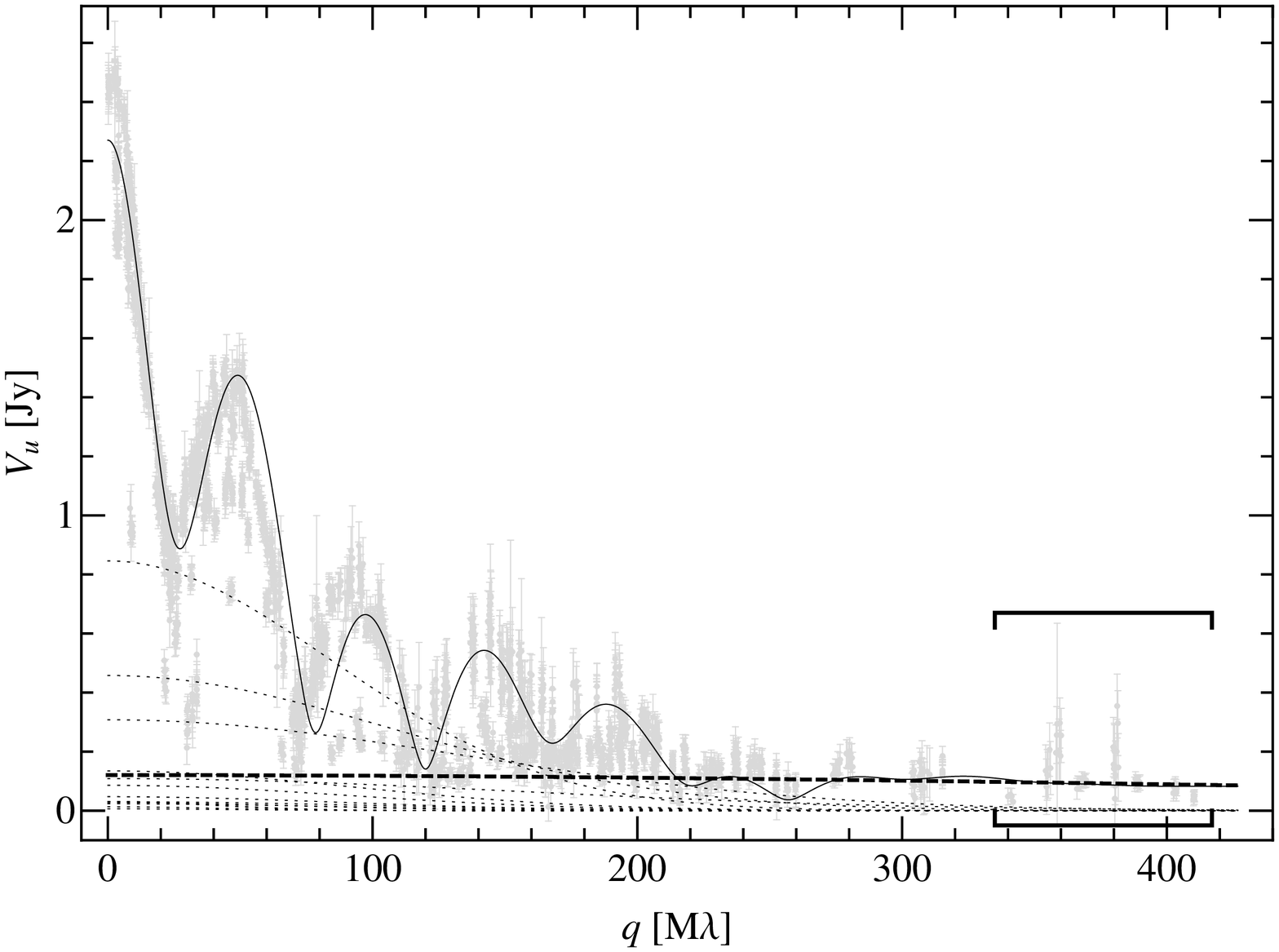}
\vskip-0.36\textwidth
\hbox to 0.465\textwidth{
\hfil
\includegraphics[height=0.25\textwidth,angle=-90,bb=67 111 496 684,clip=true]{fig02d.eps}
}
\vskip 0.172\textwidth
\caption{Comparison of the visibility amplitude distribution and
  Gaussian modelfit representation of the compact structure in a
  compact structure-dominated object (top, a 22 GHz VLBA observation
  of the quasar 3C\,345 made on 23/08/1999) and an extended
  structure-dominated object (bottom, a 15 GHz VLBA observation of the
  radio galaxy NGC\,1052 from MOJAVE Survey, made on 16/12/1995;
  \citeauthor{kellermann+2004} \citeyear{kellermann+2004}).  The
  visibility amplitude distributions are projected onto a P.A. of
  -104$^\circ$ (3C\,345) and -110$^\circ$ (NGC\,1052) for
  illustration purposes, enabling a representation by a single,
  composite modelfit rather than by fits to individual baselines. The
  insets show Gaussian modelfit images of the source structure, with
  component locations and sizes indicated. The resulting fits of the
  visibility amplitudes are shown by the solid curves. The dotted
  curves show visibility representations of the individual modelfit
  components. The thick dashed lines correspond to the respective
  visibility responses from the most compact ``core'' component that
  has the highest brightness temperature. Braces indicate the spatial
  frequency ({\em uv} distance) ranges in which the contribution from
  the core component dominates the measured visibility amplitude
  distribution (the {\em uv} ranges are measured in units of
  M$\lambda$, where $\lambda$ is the observing
  wavelength). These spatial frequency ranges are best suited to directly estimate the brightness temperature from the visibility
  measurements.}
\label{fg:visexample}
\end{figure}

Equation~\ref{eq:tbV0} provides the lowest value of $T_\mathrm{b}$ for
$V_0 = e\,V_\mathrm{q}$ (see Fig.~\ref{fg:VoVu}), which yields an
estimate of the {\em minimum brightness temperature} supported by the
measured visibility amplitude $V_\mathrm{q}$:
\begin{equation}
T_\mathrm{b,min} = \frac{\pi\,e}{2 k}\, B^2\, V_\mathrm{q}\,
\approx \, 3.09 \left(\frac{B}{\mathrm{km}}\right)^{2} \left(\frac{V_\mathrm{q}}
{\mathrm{mJy}}\right)\, \mathrm{[K]}. 
\label{eq:tbmin}
\end{equation}
This expression describes the {\em absolute minimum} of the
brightness temperature that can be obtained from the measured
visibility amplitude $V_\mathrm{q}$ under the assumption that the
brightness distribution is well approximated by a circular
Gaussian. The lowest brightness
temperature is realized for any visibility distribution that has an
inflection point. The respective expressions of $T_\mathrm{b,min}$ for
other patterns of brightness distributions are listed in Appendix~A.

\subsection{Maximum measurable brightness temperature}

The expression for $T_\mathrm{b,min}$ given by Eq.~\ref{eq:tbmin}
is independent of $V_0$, while estimating the {\em maximum} brightness
temperature will necessarily require knowledge, or at least a
reasonable assumption, of $V_0$. With the latter, it should be kept in
mind that if only a limit on $V_0$ can be assumed, the nature of the
resulting estimate of $T_\mathrm{b}$ depends on the ratio of
$V_0/V_\mathrm{q}$. The maximum $T_\mathrm{b}$ can only be derived
from {\em upper} limits on $V_0$ for $V_0 > e\, V_\mathrm{q}$ and from
{\em lower} limits on $V_0$ for $V_0 < e\, V_\mathrm{q}$. In the
opposite cases, the combination of $V_0$ and $V_\mathrm{q}$ yields an
estimate of the {\em minimum} brightness temperature.

The zero-spacing visibility $V_0$ is often approximated by the total
flux density, $S_\mathrm{tot}$, measured at a single receiving
element. Generally, this is a poor approximation because
$S_\mathrm{tot}$ contains contributions from all angular scales. A
better constraint on $V_0$ is provided by the flux density of a
respective Gaussian component if the data warrant reliable Gaussian
decomposition. For extremely poor coverages of the Fourier domain, this is
not the case, and no satisfactory estimate of $V_0$ can be made.

In this situation, a lower limit of $V_0 = V_\mathrm{q} +
\sigma_\mathrm{q}$ can be adopted.  This limit effectively corresponds to requiring that $V_\mathrm{q}$ probes a structural detail that is
marginally resolved (we recall that $V_\mathrm{q} \equiv const$ results
from the Fourier transform of a point source). This assumption is well
justified for visibility measurements made at sufficiently long
baselines, where the visibility amplitude is dominated by the most
compact structure observed in the target object. It can furthermore
be verified through the observed absence of amplitude beating at slightly
shorter baselines (with the latter reflecting the presence of multiple
compact emitting regions in the object).  Examples of visibility
(baseline) ranges dominated by contributions from the most compact
structures are shown in Fig.~\ref{fg:visexample} for two radio sources
(NGC\,1052 and 3C\,345) that represent the typical cases of a compact
radio source with and without a strong contribution from extended
emission.

The requirement of marginal resolution of the observed structure
implies that its size should be larger than
\begin{equation}
\theta_\mathrm{lim} = \frac{2 \sqrt{\ln 2}}{\pi}\frac{\lambda}{B}\sqrt{\ln\frac{V_\mathrm{q}+\sigma_\mathrm{q}}{V_\mathrm{q}}}\,.
\end{equation}
Correspondingly, the brightness temperature of this feature should not
exceed the limit of
\begin{eqnarray}\nonumber
T_\mathrm{b,lim} &=& \frac{\pi B^2\, (V_\mathrm{q}+\sigma_\mathrm{q})}{2 k} \left[\ln \frac{V_\mathrm{q} + \sigma_\mathrm{q}}{V_\mathrm{q}}\right]^{-1}\\
& = & 1.14 \left(\frac{V_\mathrm{q}+\sigma_\mathrm{q}}{\mathrm{mJy}}\right)\left(\frac{B}{\mathrm{km}}\right)^{2} \left(\ln \frac{V_\mathrm{q}+\sigma_\mathrm{q}}{V_\mathrm{q}}\right)^{-1}\,\mathrm{[K]}\,.
\label{eq:tblim}
\end{eqnarray}

\begin{figure}[t!]
\begin{center}
\includegraphics[width=0.48\textwidth]{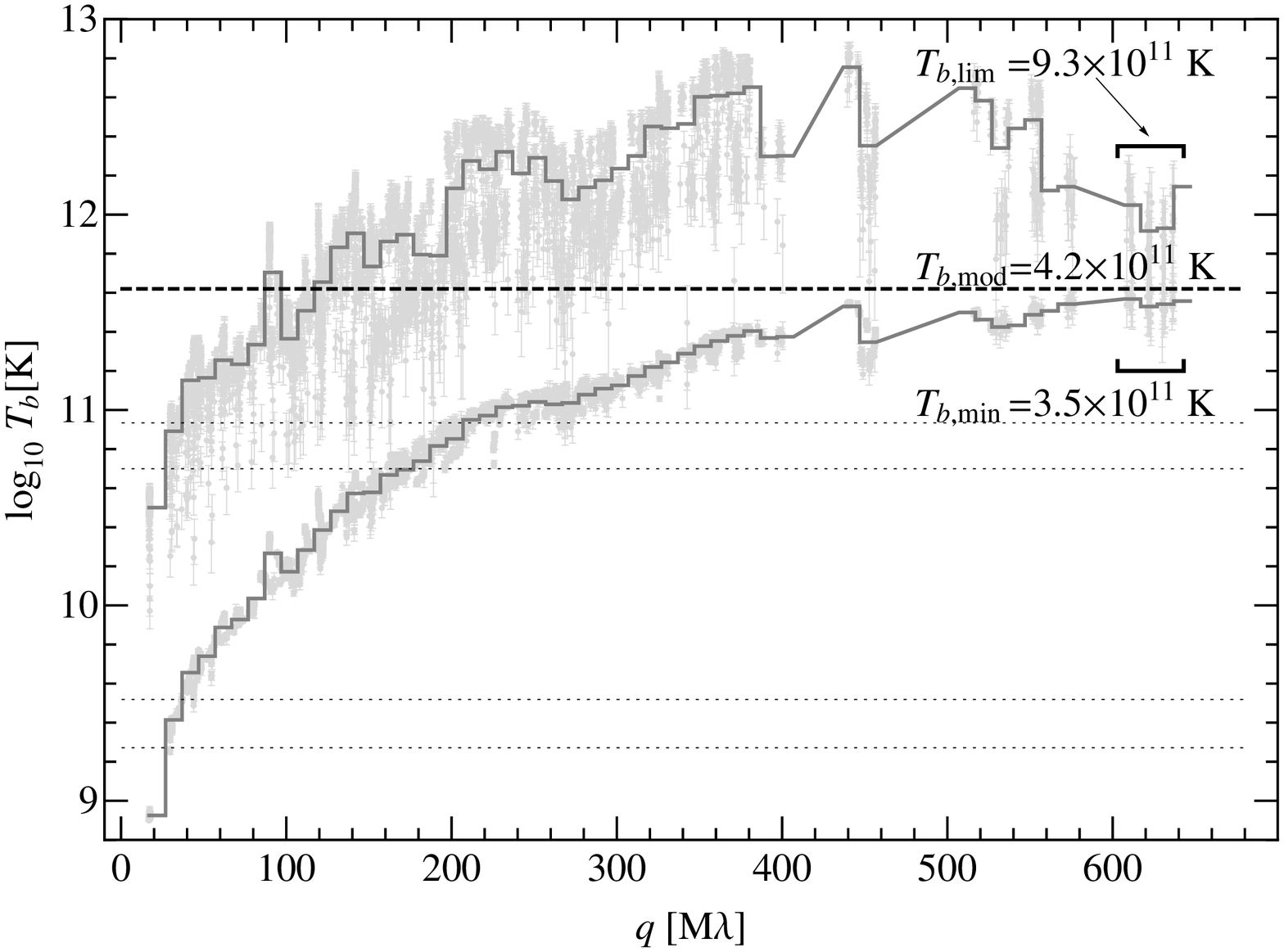}
\includegraphics[width=0.48\textwidth]{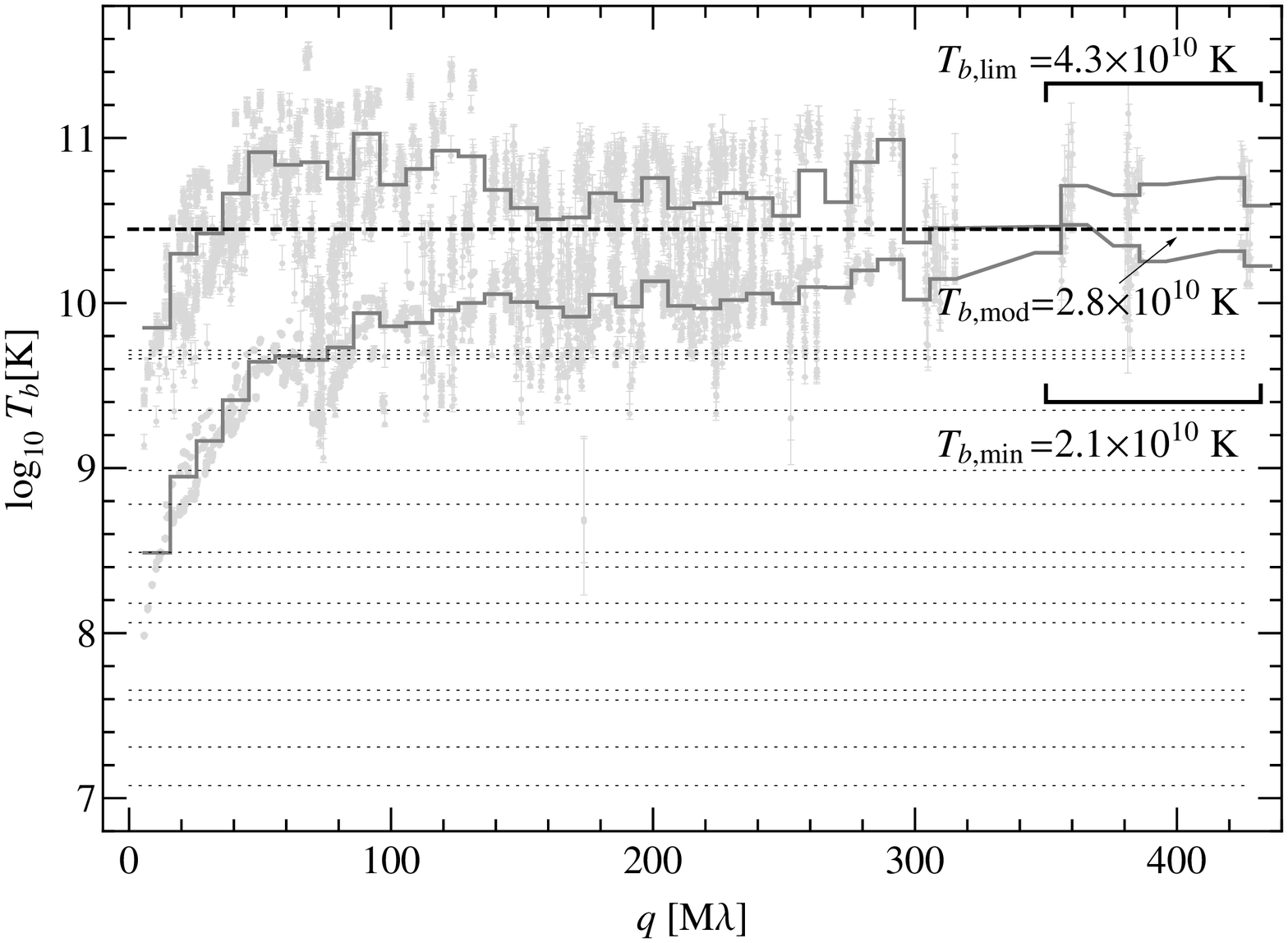}
\end{center}
\caption{Limiting, $T_\mathrm{b,lim}$ (top row of datapoints) and
  minimum, $T_\mathrm{b,min}$ (bottom row of datapoints), brightness temperature
  estimated directly from visibility data for 3C\,345 (top panel) and NGC\,1052 (bottom panel),
  plotted against the spatial frequency $q$ ({\em uv} radius).  The
  estimated values are compared with the brightness temperatures of the
  Gaussian model fit components (dotted lines for the jet components,
  thick dashed line for the core component) used to describe the
  source structure as shown in Fig.~2. Running stairs show the
  respective rows of the brightness temperature averaged within radial
  bins of 10\,M$\lambda$ in extent. Both the original and the averaged
  rows of the brightness temperature indicate that at $q\gtrsim
  150$\,M$\lambda$, the interval $[T_\mathrm{b,min},T_\mathrm{b,lim}]$
  provides a good bracketing for the maximum brightness temperature in
  each of the two objects. Braces indicate the conservative
  ranges of spatial frequency $q$ identified as ranges dominated by the most compact
part of the source structure. Averages of $T_\mathrm{b,min}$ and $T_\mathrm{b,lim}$ made over these ranges constrain the respective $T_\mathrm{b,mod}$ estimates well.}
\label{fg:tbexample}
\end{figure}

\noindent
For visibilities with a signal-to-noise ratio
$V_\mathrm{q}/\sigma_\mathrm{q} > e-1$, the estimate of
$T_\mathrm{b,lim}$ provides the highest brightness
temperature that can be obtained from the measured visibility
amplitude and its error while requiring that the respective brightness
distribution is {\em a)} circularly Gaussian, and {\em b)} marginally
resolved by the measured visibility. Expressions for
$T_\mathrm{b,lim}$ derived for several other brightness
distribution patterns are given in Appendix~A.

As measurements of $V_\mathrm{q}$ and $\sigma_\mathrm{q}$ are made
over extended time intervals, they correspond to averaging the
visibility function over finite ranges of spatial frequencies ($\Delta
q,\,\Delta \psi$), with $\psi$ describing the positional angle of
$V_\mathrm{q}$ in the Fourier plane. In this case, the estimate
provided by Eq.~\ref{eq:tblim} holds for as long as the condition
$V_\mathrm{q,\psi} \approx const$ is satisfied over the given measurement interval [$\Delta q,\,\Delta \psi$]. This condition effectively requires
that the measured $V_\mathrm{q}$ is dominated by a
contribution from a single emitting region that is marginally resolved at the
spatial frequency $q$.

Equations \ref{eq:tbmin} and \ref{eq:tblim} provide a robust bracketing
for the brightness temperature obtained from interferometric measurements that
have a limited sampling of the visibility distribution of the
target. This can be demonstrated by applying these equations to every
visibility of the VLBI datasets used in the examples shown in
Fig.~\ref{fg:visexample}. Results of this application are presented in
Fig.~\ref{fg:tbexample}, where the visibility-based estimates of
$T_\mathrm{b,min}$ and $T_\mathrm{b,lim}$ are plotted against the
spatial frequency ({\em uv} radius) of the respective
visibilities. These estimates can be compared with the brightness
temperature, $T_\mathrm{b,mod}$ estimated from the model fit
parameters of the ``core'' component. This comparison indicates that
at {\em uv} radii $\gtrsim 150$\,M$\lambda$, the interval
$[T_\mathrm{b,min},T_\mathrm{b,lim}]$ represents a reasonably good
bracketing for the expected maximum brightness temperature. This
substantially exceeds the conservative expectations for the {\em
  uv} ranges (indicated by braces in Fig.~\ref{fg:tbexample}) suitable
for estimating the brightness temperature. Within these ranges,
the average $T_\mathrm{b,lim}$ is only marginally (factors of 1.5 and
2.2) higher than the $T_\mathrm{b,mod}$ obtained from modelfits. Hence,
$T_\mathrm{b,lim}$ estimates made at long baselines can
constrain  the maximum brightness temperature well in both
core-dominated and jet-dominated compact radio sources.

\begin{figure}[t!]
\begin{center}
\includegraphics[width=0.48\textwidth]{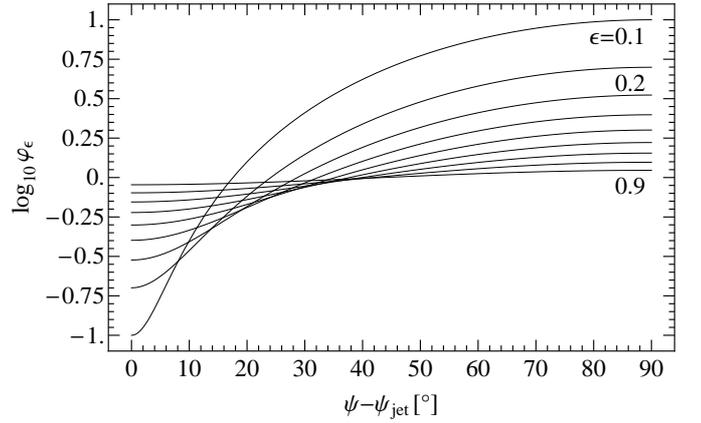}
\end{center}
\caption{Brightness temperature correction to account for the
elongation
  of the emitting region. The elongated region is described by an
  elliptical Gaussian with an axial ratio $\epsilon$. The corrections
  are plotted against the difference between the position angle of a
  visibility measurement and the position angle of the major axis of
  the Gaussian. Different curves correspond to the corrections for values
  of $\epsilon$ taken in steps of 0.1 between $\epsilon=0.1$ and
  $\epsilon= 0.9$.}
\label{fg:corrfac}
\end{figure}

\subsection{Corrections for elongation of the emitting region}

The estimates of $T_\mathrm{b,min}$ and $T_\mathrm{b,lim}$ can furthermore be refined if the potential elongation of the emitting region is
considered.  If visibility measurements are made over a narrow range
of position angle $\psi$ in the Fourier plane, this elongation may
bias individual estimates of the brightness temperature and introduce
scatter in the statistics obtained from object samples. Gaussian
modeling of fine structure in the objects from the MOJAVE sample
\citep{kovalev+2005} indicates that the core components are well
described by elliptical Gaussian patterns, with an average elongation
(minor to major axis ratio, $\epsilon =
\theta_\mathrm{min}/\theta_\mathrm{maj}$) of $0.4 \pm 0.2$. For visibility
measurements made at a random position angle $\psi$ (hence oriented
randomly with respect to the source elongation), the resulting scatter
of $T_\mathrm{b}$ estimates can approach one order of magnitude. This
adverse effect of the source elongation can be taken into account in
objects with a known elongation and position angle of the compact
structure.

The compact core region can be approximated by an elliptical Gaussian
with an axial ratio, $\epsilon$, and a position angle of the major
axis, $\psi_\mathrm{jet}$. The angular size estimate,
$\theta_\mathrm{r}$, obtained under the assumption of a circular
Gaussian brightness distribution ({\em e.g.}, with
Eq.~\ref{eq:thetar}) can then be related to the major and minor
axes of the elliptical Gaussian,
\begin{eqnarray*}
\theta_\mathrm{maj}& = &\theta_\mathrm{r} /\sqrt{\sin^2\zeta + \epsilon^2 \cos^2\zeta}\,,\\
\theta_\mathrm{min}&  = & \theta_\mathrm{r}/ \sqrt{\cos^2\zeta + \epsilon^{-2} \sin^2\zeta} \equiv \theta_\mathrm{maj}\,\epsilon\,,
\end{eqnarray*}
where $\zeta = \psi - \psi_\mathrm{jet}$ describes the difference
between the visibility position angle and that of
the major axis of the elliptical Gaussian component (hence $\theta_\mathrm{r} =
\theta_\mathrm{min}$ for $\zeta=0^{\circ}$ and
$\theta_\mathrm{r}=\theta_\mathrm{maj}$ for $\zeta = 90^{\circ}$).  For the
brightness temperature estimates, this results in a multiplicative
correction factor
\begin{equation}
\varphi_{\epsilon} = \theta_\mathrm{r}^2/(\theta_\mathrm{maj}\,\theta_\mathrm{min}) = \epsilon\, \cos^2\zeta + \epsilon^{-1}\, \sin^2\zeta
\label{eq:corrfac}
\end{equation}
that should be applied to $T_\mathrm{b,min}$ and $T_\mathrm{b,lim}$
given by Eqs.~\ref{eq:tbmin} and 5. The magnitude of this
correction is on the order of $1/\epsilon^2$ over the full range of
values of $\zeta$ (see Fig.~\ref{fg:corrfac}). Applying this
correction may be particularly useful to analyze
space VLBI measurements made with RadioAstron at baselines in excess
of ten Earth diameters (hence falling within a range of $\Delta \psi
\lesssim 6^{\circ}$). Other potential applications include
snapshot VLBI measurements made at millimeter
wavelengths \citep[{\em e.g.},][]{doeleman+2012,petrov+2012,lee+2013}.

\begin{figure}[t!]
\begin{center}
\includegraphics[width=0.48\textwidth]{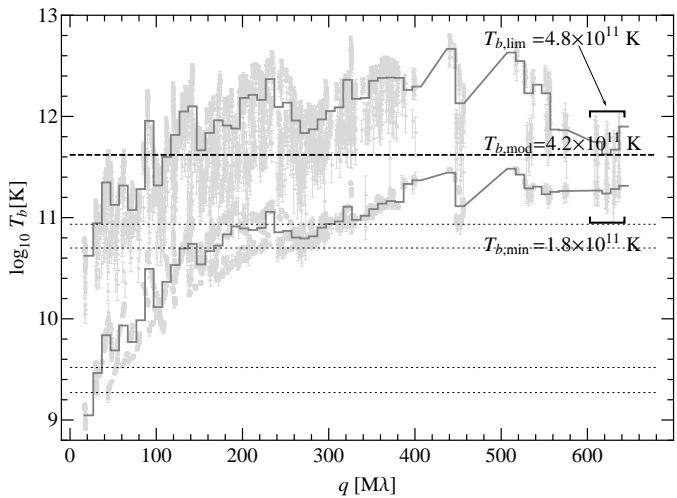}
\end{center}
\caption{The limiting, $T_\mathrm{b,lim}$ (top run) and
  minimum, $T_\mathrm{b,min}$ (bottom run), brightness temperature
  estimated directly from visibility data for 3C\,345 and corrected
  for an assumed ellipticity of the emitting region, with $\epsilon =
  0.5$ (based on MOJAVE model fitting reported in
  \citeauthor{kovalev+2005} \citeyear{kovalev+2005}). and
  $\psi_\mathrm{jet} = -104^{\circ}$. The limiting brightness
  temperature of $4.8\times 10^{11}$\,K, obtained from the visibilities
  at the longest baselines ($B>600\,\mathrm{M}\lambda$), agrees well
  with the estimate based on model fitting the source structure.}
\label{fg:tbcorr}
\end{figure}

The effect of correcting for the core elongation is illustrated in
Fig.~\ref{fg:tbcorr}, where the correcting factor $\varphi_\epsilon$ is
applied to brightness temperature estimates made from the visibility
data on 3C\,345 presented in
Figs.~\ref{fg:visexample}-\ref{fg:tbexample}. The corrections are
derived for an axial ratio $\epsilon = 0.5$ (weighted average of the
model-fitting results in \citeauthor{kovalev+2005}
\citeyear{kovalev+2005}) and a jet position angle $\psi_\mathrm{jet} =
-104^{\circ}$ inferred from the source structure shown in
Fig.~\ref{fg:visexample}. The increased scatter at shorter baselines
shows the effect of contributions from larger structures that are incorrectly described by the adopted values of $\phi_\mathrm{jet}$ and,
in particular, $\epsilon$. However, at the longest baselines,
correcting for the core elongation clearly improves the
$T_\mathrm{b,lim}$ estimate and brings it well within the errors of
the modelfit-based estimate. The same correction applied to NGC\,1052
(with $\epsilon = 0.4$ and $\psi_\mathrm{jet} = -110^{\circ}$) results
in corrected $T_\mathrm{b,min} = 1.1 \times 10^{10}$\,K and
$T_\mathrm{b,lim} = 2.4 \times 10^{10}$\,K, with the latter value
falling very close to $T_\mathrm{b,mod} = 2.8\times 10^{10}$\,K
estimated from the modelfit. Both these results indicate that
$T_\mathrm{b,lim}$ can be improved by correcting for the elongation of
the core region.

\section{Discussion}
\label{sc:discussion}

Applying\ the visibility-based brightness
temperature estimates to VLBI data on 3C\,345 and NGC\,1052 has demonstrated
that  the method is reliable in two particular cases. A more extended
testing of the method can be performed on a statistical basis by
applying it to visibility data from large VLBI survey programs aimed
at measuring and analyzing the brightness temperature distribution in
samples of compact radio sources.  The analysis of fine scale
structure in the 15\,GHz ($\lambda_\mathrm{obs} = 2$\,cm) MOJAVE
sample \citep{kovalev+2005} and the results of brightness temperature
measurements from the 86 GHz ($\lambda_\mathrm{obs} = 3$\,mm) CMVA
survey \cite{lee+2008} offer statistically suitable samples for such
tests.  The MOJAVE analysis was based on elliptical Gaussian model
fits of the core region. The 86 GHz data were fitted by circular
Gaussian components. Resolution criteria were applied to the data from
both surveys to constrain the core components with degenerate size
parameters (circular diameter or one of the two axes of elliptical
Gaussian) obtained from the model fitting.

\begin{figure}[t!]
\includegraphics[width=0.48\textwidth]{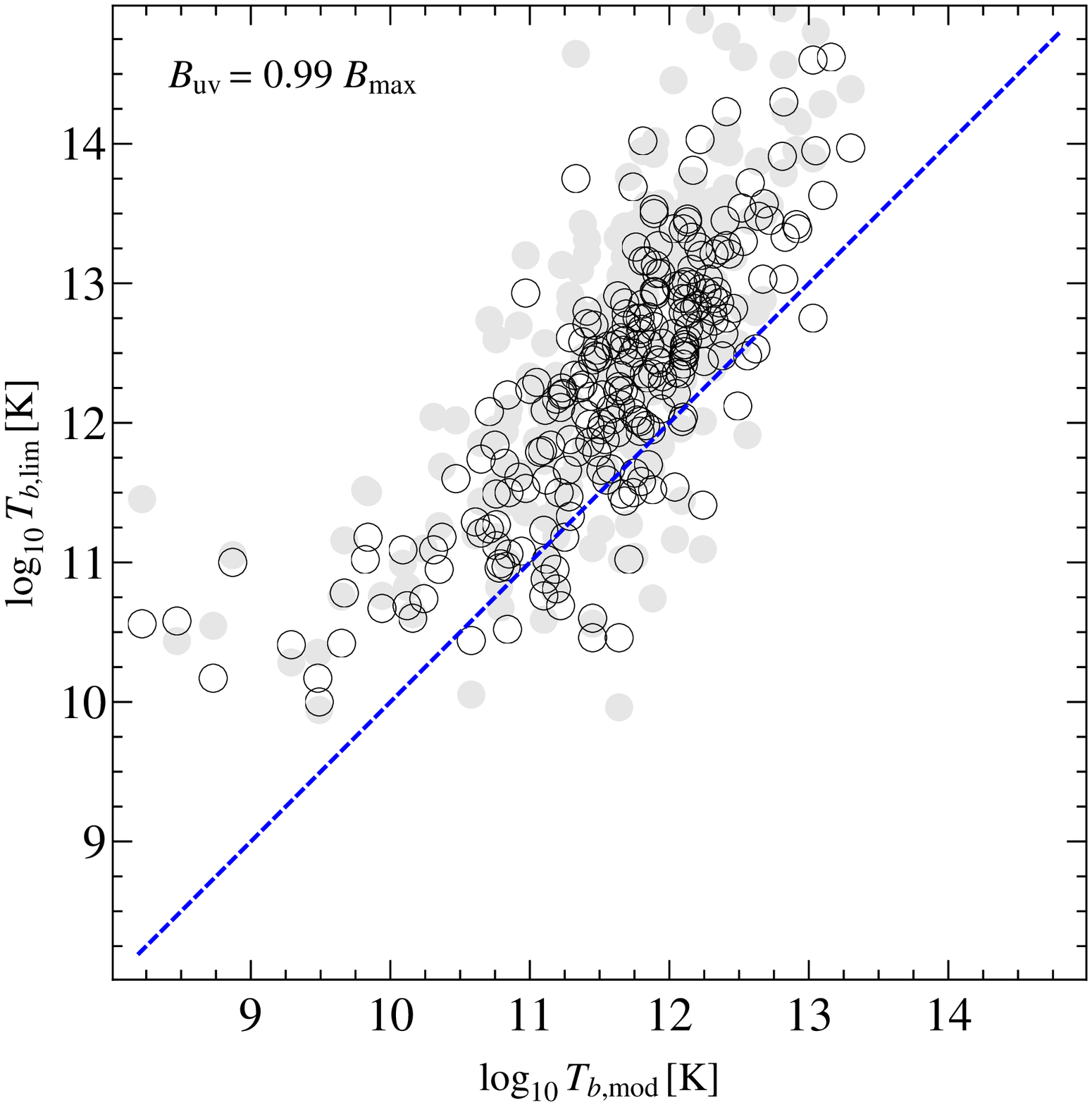}
\vskip-0.23\textwidth
\hbox to 0.47\textwidth{
\hfil
\includegraphics[height=0.16\textwidth]{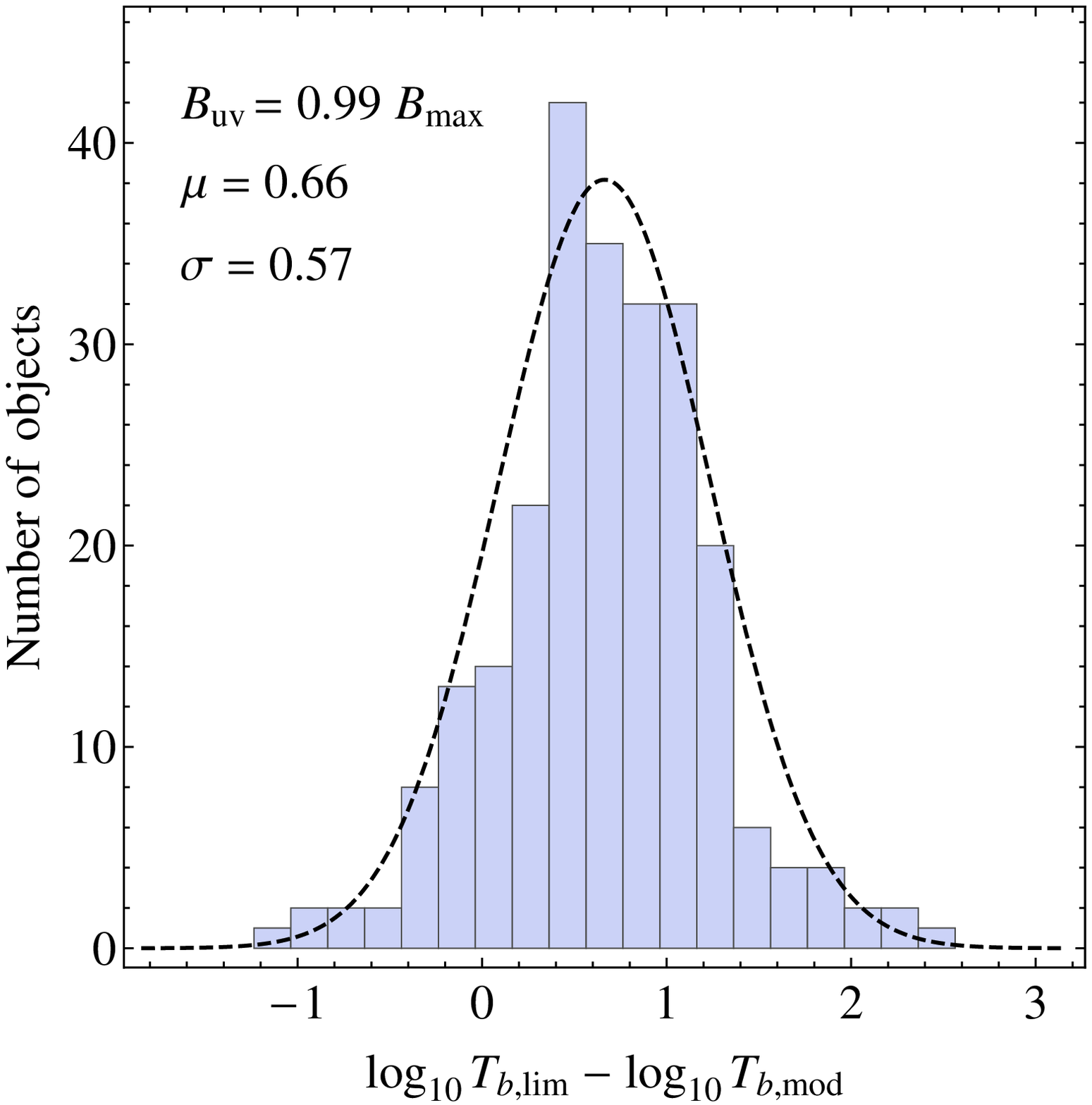}
}
\vskip 0.08\textwidth
\caption{Comparison of $T_\mathrm{b,lim}$ (circular Gaussian
  approximation) and $T_\mathrm{b,mod}$ estimates obtained from the
  MOJAVE data (open circles). Gray circles illustrate the effect of
  correcting $T_\mathrm{b,lim}$ for the putative elongation of the
  core region. The dashed line marks the one-to-one correspondence
  between the two estimates. For each object, the $T_\mathrm{b,lim}$
  is estimated from MOJAVE data at $R_\mathrm{uv}\ge
  0.99\,B_\mathrm{max}$ to restrict the visibility
  information to the most compact structures. The resulting
  $T_\mathrm{b,lim}$ are on average 4.6 times higher than
  $T_\mathrm{b,mod}$. The residual logarithmic distribution of the
  $T_\mathrm{b,lim}/T_\mathrm{b,mod}$ ratio (inset) can be
  approximated by the Gaussian PDF with $\mu =0.66$ and $\sigma =
  0.57$. At $T_\mathrm{b,mod}\lesssim 10^{10}$\,K the
  $T_\mathrm{b,lim}$ may be biased by large-scale structure
  contributions in strongly jet-dominated objects. }
\label{fg:mojave99}
\end{figure}

\subsection{Preparation of the visibility data}

To evaluate the performance of visibility-based brightness temperature
limits on a self-consistent statistical basis, visibility data for
each source from these two programs (244 objects in the MOJAVE sample
and 123 observations of 109 individual objects in the CMVA sample) were
radially and azimuthally averaged within a small annulus,
$(B_\mathrm{uv}, B_\mathrm{max})$ in the {\em uv} plane. The purpose
of clipping the data is to reduce the adverse effect of including
shorter baselines that are dominated by extended structure. The upper limit of
the annulus is determined by the longest baseline $B_\mathrm{max}$
found in the data for each individual object.  The MOJAVE datasets were
clipped at $B_\mathrm{uv}=0.99\,B_\mathrm{max}$ ({\em i.e.}, in an
annulus located within 1\% of $B_\mathrm{max}$). The CMVA survey data,
with substantially fewer visibilities per target object,
were clipped at $B_\mathrm{uv} =0.9\,B_\mathrm{max}$. The resulting
average {\em uv} distances and visibility position angles in the
clipped data are $435 \pm 27$ M$\lambda$ and $-88^{\circ}\pm
6^{\circ}$ for the MOJAVE data and $2270 \pm 660$ M$\lambda$ and
$-78^{\circ} \pm 20^{\circ}$ for the CMVA data. In each case, the
  fraction of visibilities selected is $\sim 1/N_\mathrm{bas}$, where
  $N_\mathrm{bas}$ is the number of baselines in a dataset.
Following the conclusions obtained from analyzing the data on 3C\,345
and NGC\,1052, these {\em uv} distances should be sufficiently long
to reduce the visibility contamination by large-scale structures to
negligible levels.  The $T_\mathrm{b,lim}$ estimates were therefore
calculated for each source from all of the visibilities clipped and
averaged within the respective annuli. The circular Gaussian
approximation was used in these calculation.

\subsection{Results from the MOJAVE data}
\label{sc:res_mojave}

For the MOJAVE data, the resulting estimates of
$T_\mathrm{b,lim}$ are compared in Fig.~\ref{fg:mojave99} with the
$T_\mathrm{b,mod}$ estimates obtained in \cite{kovalev+2005} for each
source at the same observing epoch.  The two estimates agree
reasonably well, with $T_\mathrm{b,lim}$ being on average 4.6 times
higher than the respective $T_\mathrm{b,mod}$. This is similar to the
$T_\mathrm{b,lim}/T_\mathrm{b,mod}$ ratios measured in 3C\,345 and
NGC\,1052. Overall, the $T_\mathrm{b,lim}$ estimates provide a
reasonable upper bound on the brightness temperature, with the scatter
in the estimates limited to about half a decade over nearly five
orders of magnitude in brightness temperature. 

Correction for the putative elongation of the core region has been
attempted for the MOJAVE data, based on the elliptical model fits and
measured position angles of the jet. The elongation was calculated as
the ratio of the minor to major axes of the elliptical Gaussian
component describing the core.  Two options were tried for the jet
position angle: a) the position angle of the major axis of the
Gaussian component, and b) the average position angle of the jet as reported
in \cite{kovalev+2005}. The results of applying the elongation
correction with the former option are shown in
Fig.~\ref{fg:mojave99}. Neither of the two corrections has improved
the correlation for either the average
$T_\mathrm{b,lim}/T_\mathrm{b,mod}$ ratio or the spread of the
residuals. 

\begin{figure}[t!]
\includegraphics[width=0.48\textwidth]{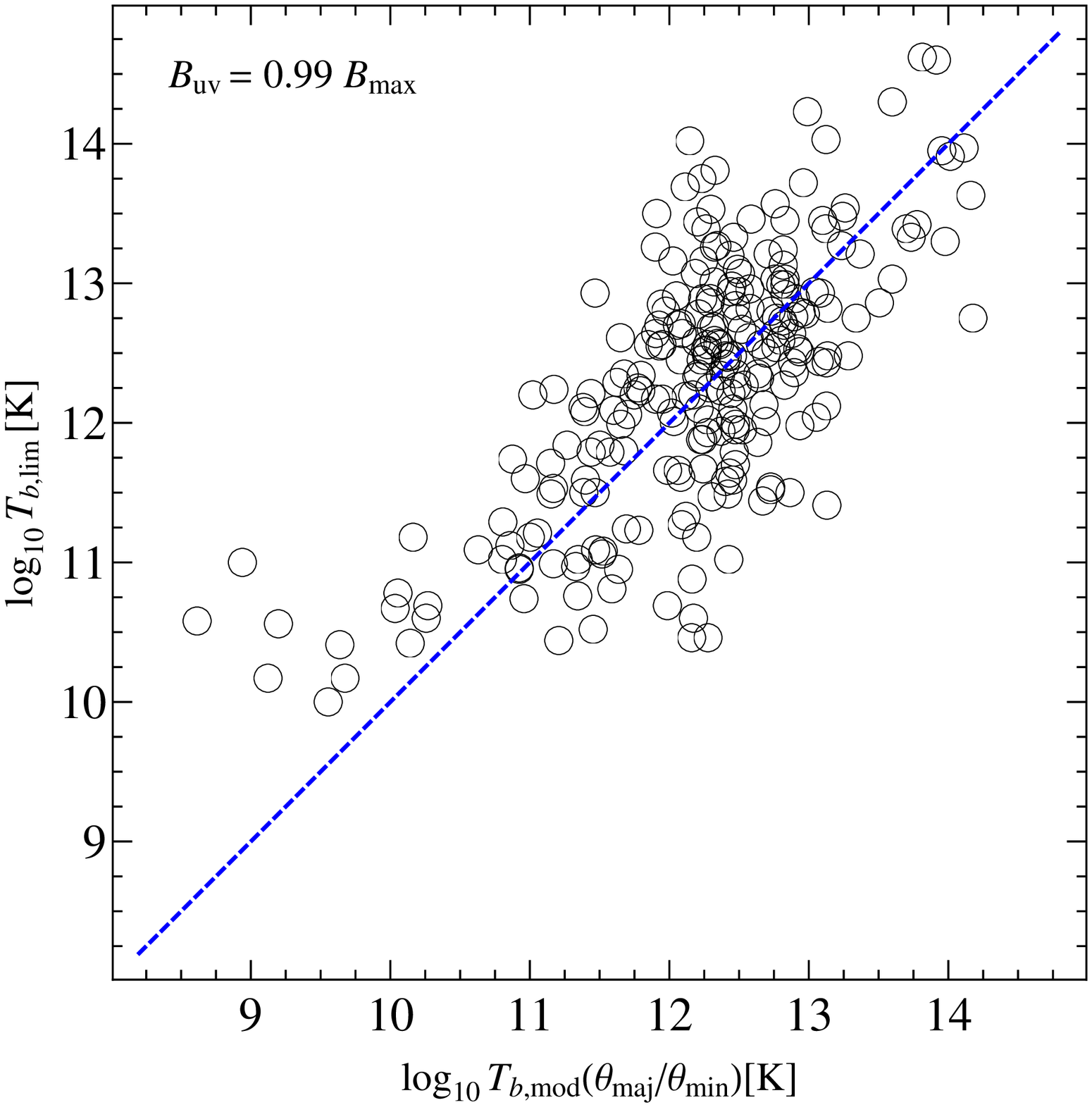}
\vskip-0.23\textwidth
\hbox to 0.47\textwidth{
\hfil
\includegraphics[height=0.16\textwidth]{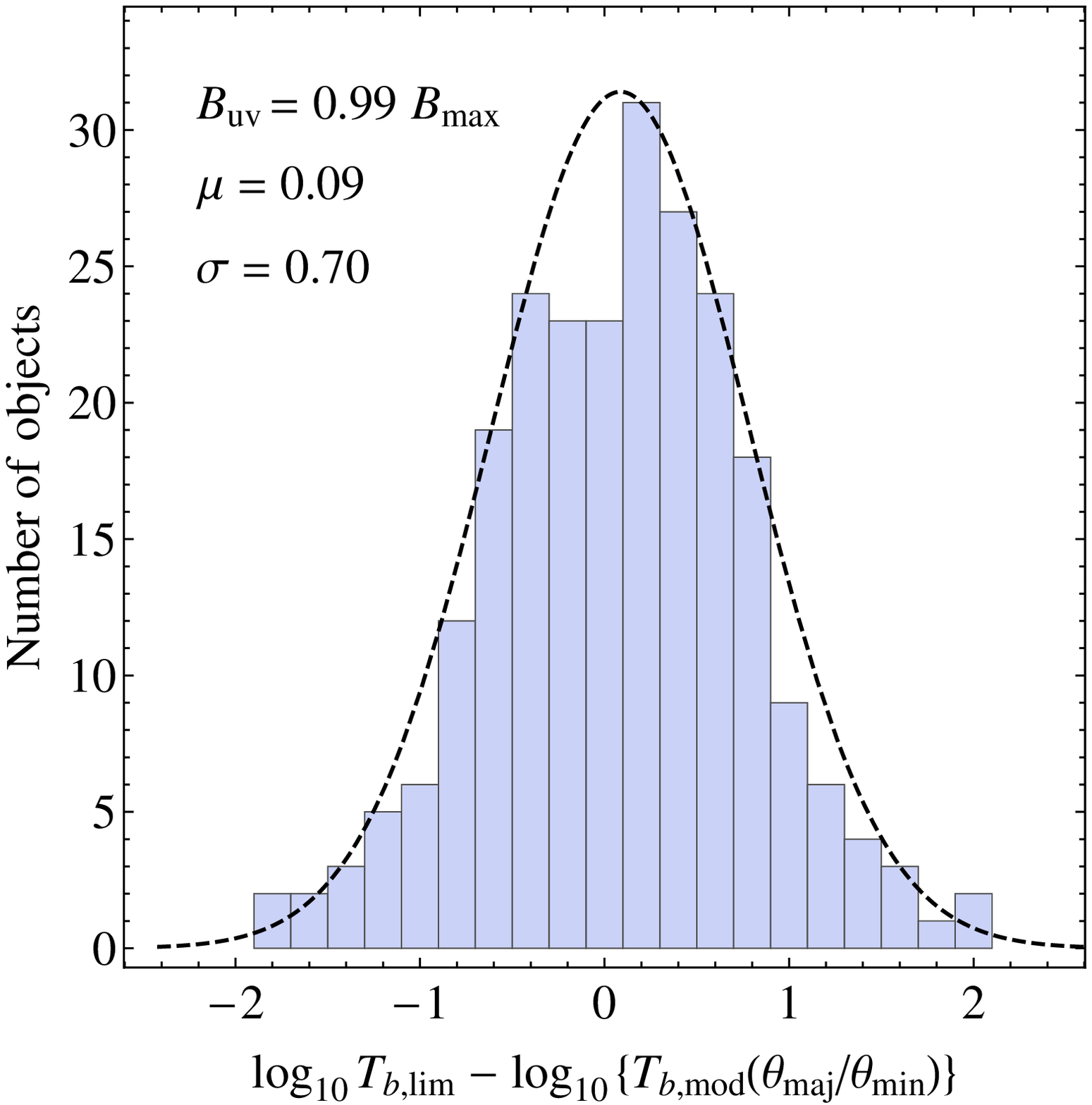}
}
\vskip 0.08\textwidth
\caption{Correlation between $T_\mathrm{b,lim}$ and
  $T_\mathrm{b,mod}(\theta_\mathrm{maj}/\theta_\mathrm{min})$
  expressing the brightness temperature obtained assuming that
  the characteristic size of the brightest region is determined by the
  jet transverse dimension as given by the minor axis of the elliptical
  Gaussian fit. The average $T_\mathrm{b,lim}$ is is only 20\% higher
  than the respective corrected values of $T_\mathrm{b,mod}$ , and the residual
  logarithmic distribution of the $T_\mathrm{b,lim}/T_\mathrm{b,mod}$
  ratio (inset) can be approximated by the Gaussian PDF with $\mu
  =0.09$ and $\sigma = 0.70$.}
\label{fg:tbbmin}
\end{figure}

\begin{figure}[t!]
\includegraphics[width=0.48\textwidth]{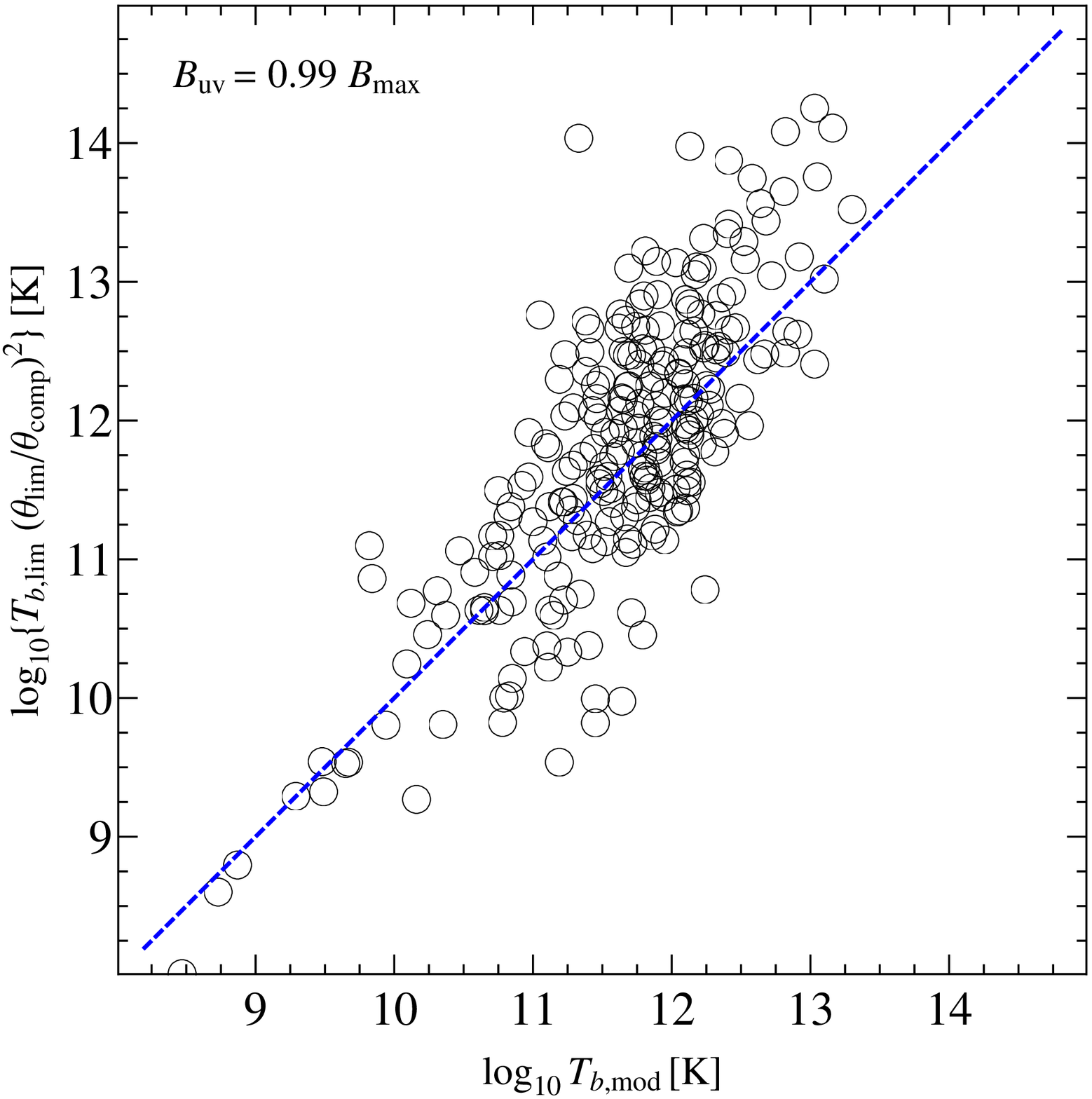}
\vskip-0.23\textwidth
\hbox to 0.47\textwidth{
\hfil
\includegraphics[height=0.16\textwidth]{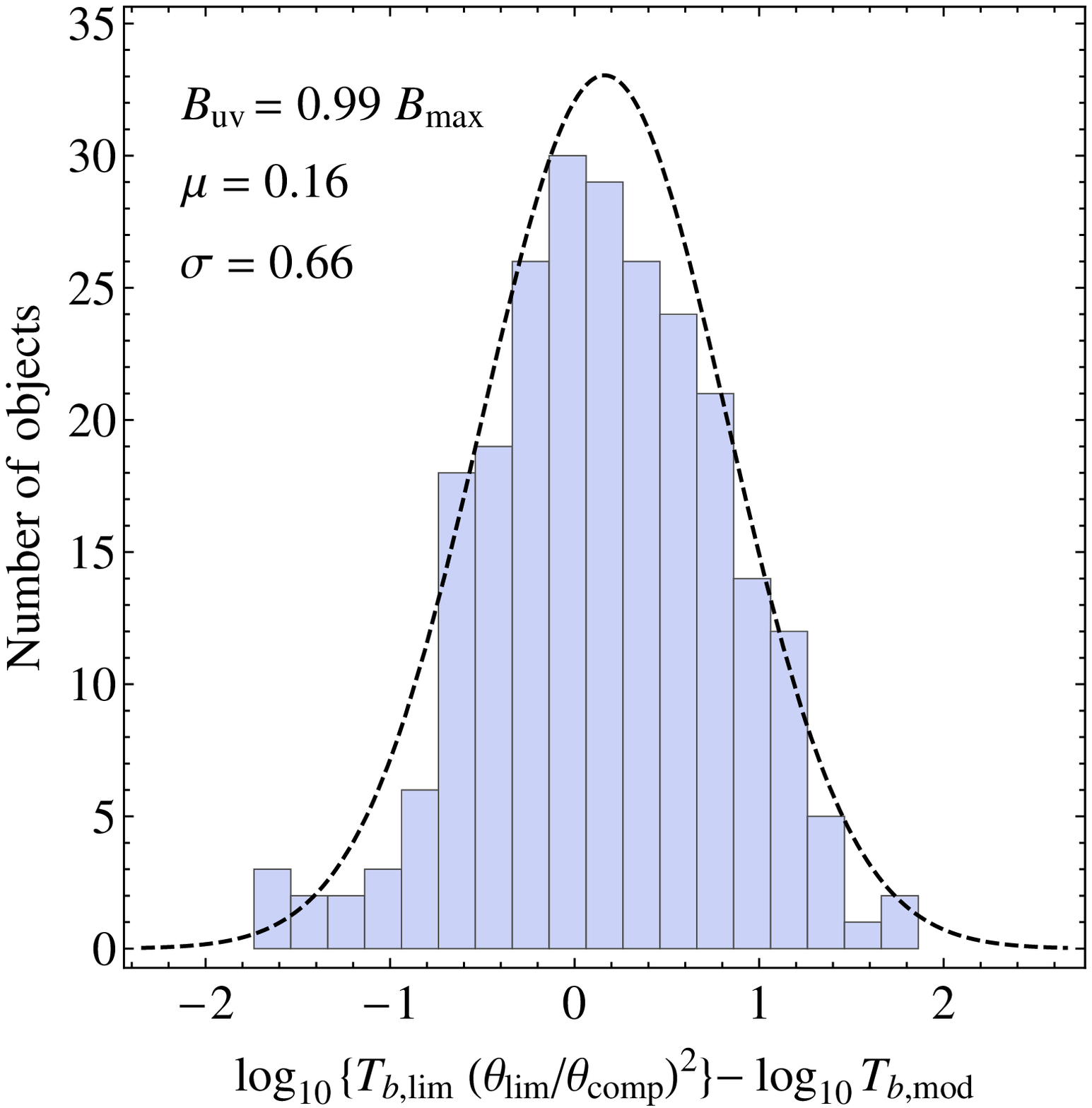}
}
\vskip 0.08\textwidth
\caption{Correction of the distribution shown in
  Fig.~\ref{fg:mojave99} for the square of the resolution factor
  $\theta_\mathrm{lim}/\theta_\mathrm{comp}$. The corrected
  $T_\mathrm{b,lim}$ distribution is statistically very close to the
  $T_\mathrm{b,mod}$ obtained from the model-fitting analysis. The
  values of $T_\mathrm{b,lim}$ are on average 40\% higher than the
  respective values of $T_\mathrm{b,mod}$. This is also demonstrated
  by the residual logarithmic distribution of the
  $T_\mathrm{b,lim}/T_\mathrm{b,mod}$ ratio (inset), which can be
  approximated by the Gaussian PDF with $\mu =0.16$ and $\sigma =
  0.66$.}
\label{fg:tbres}
\end{figure}

This result may be caused by two factors. On the one hand, it
may suggest that the elongation of the core region is not accurately reflected in the parameters of the elliptical Gaussian components,
for instance, if this region has a specific geometry such as a
conically expanding jet \citep[{\em cf}.][]{blandford1979} or a jet
pervaded with thread-like instability patterns \citep[{\em
  cf}.][]{lobanov2001}. On the other hand, the lack of improvement
achieved with the elongation correction may indicate that the highest
brightness is realized in a region that is smaller than the major
axis of the Gaussian component describing the core region. Again,
this would be the case for a quasi-stationary conical jet or a jet
dominated by the instability patterns. In each of these situations, the
observed jet brightness would be largely determined by its transverse
dimension. If this is the case, $T_\mathrm{b,lim}$ could be reconciled
with $T_\mathrm{b,mod}$ derived under the assumption that only the minor
axis, $\theta_\mathrm{min}$, of the Gaussian component is relevant for
determining the jet brightness.

This hypothesis is tested in Fig.~\ref{fg:tbbmin} by comparing
$T_\mathrm{b,lim}$ to
$T_\mathrm{b,mod}\,\theta_\mathrm{maj}/\theta_\mathrm{min}$, which has
the effect of calculating the size of the core region as
$\pi\,\theta_\mathrm{min}^2/4$. This simple correction brings the two
estimates to a very good agreement, with $T_\mathrm{b,lim}$ estimates,
as demonstrated by both the average ratio between the two estimates
and the distribution of logarithmic residuals of this ratio. The only
notable discrepancy between the two estimates is the persistently
higher $T_\mathrm{b,lim}$ values observed for the low
$T_\mathrm{b,mod} \lesssim 3\times 10^{10}$. Such brightness
temperatures are typically measured in objects in which the cores are
strongly resolved along both axes of the fitted Gaussians, hence they
may require a correction for the resolution along the minor axis of
the Gaussian as well.

Such a correction can be performed by substituting the fitted
effective size, $\theta_\mathrm{comp} =
\sqrt{\theta_\mathrm{min}\theta_\mathrm{maj}}$ of the Gaussian
component with its respective resolution limit $\theta_\mathrm{lim}$ derived in
\cite{kovalev+2005}. This
operation should account for resolving the core region along and
across the jet direction as marked by the position angle of the major
axis of the fitted Gaussian component. Applying this correction
is illustrated by Fig.~\ref{fg:tbres}, which shows an excellent agreement
between the two estimates of the brightness temperature and indeed
effectively brings the low $T_\mathrm{b,mod}$ outliers to
the main correlation trend.

The examples set by substituting the measured core size by either
$\theta_\mathrm{min}$ or $\theta_\mathrm{lim}$ indicate that the
visibility-based estimates of $T_\mathrm{b,lim}$ can be related to
brightness temperature estimates obtained from Gaussian model fitting
and that the $T_\mathrm{b,lim}$ estimates present a reliable account of
brightness temperatures measured at the {\em limit of angular
  resolution} of the respective interferometric data sets.

\begin{figure}[t!]
\includegraphics[width=0.48\textwidth]{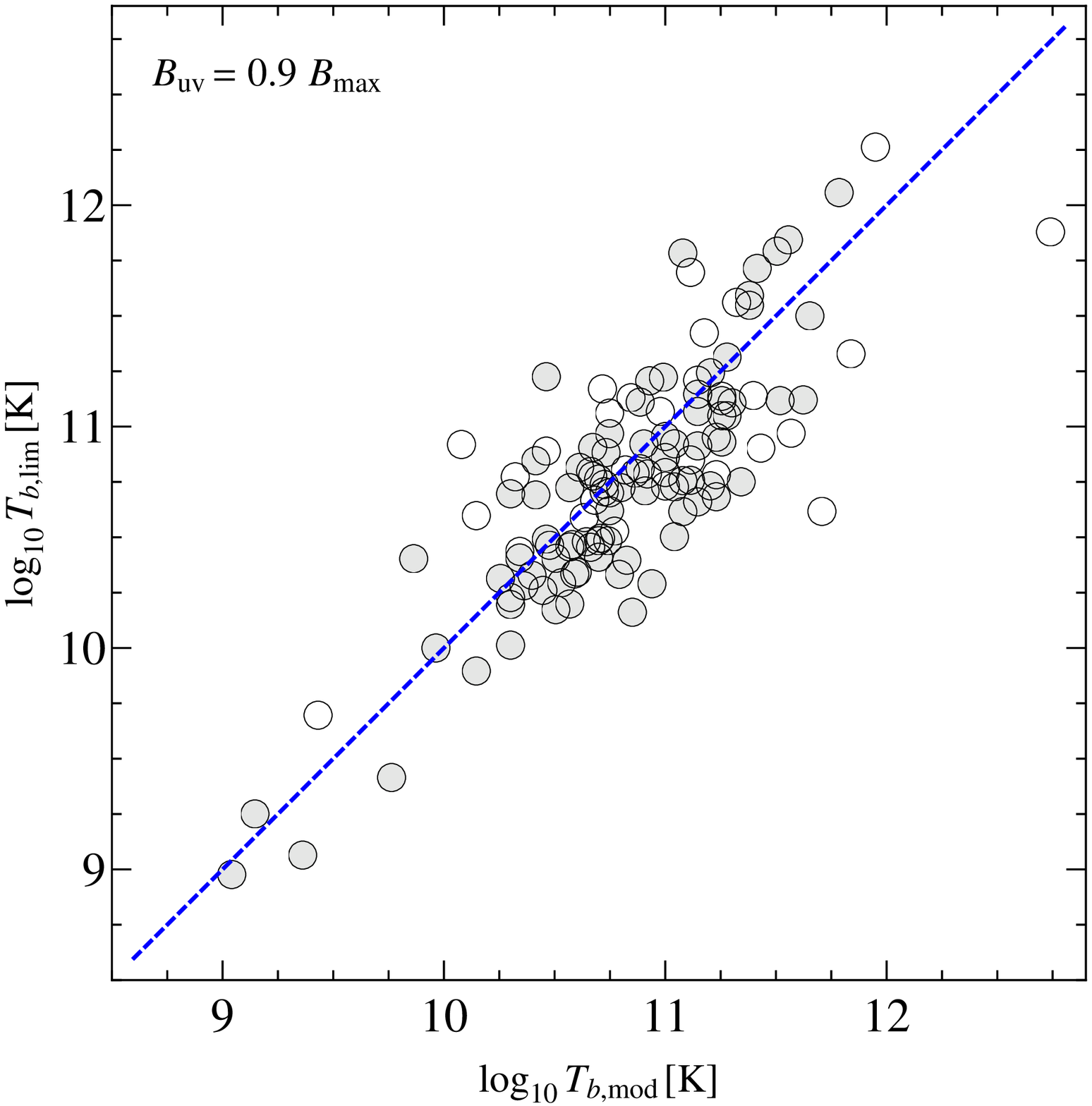}
\vskip-0.23\textwidth
\hbox to 0.47\textwidth{
\hfil
\includegraphics[height=0.16\textwidth]{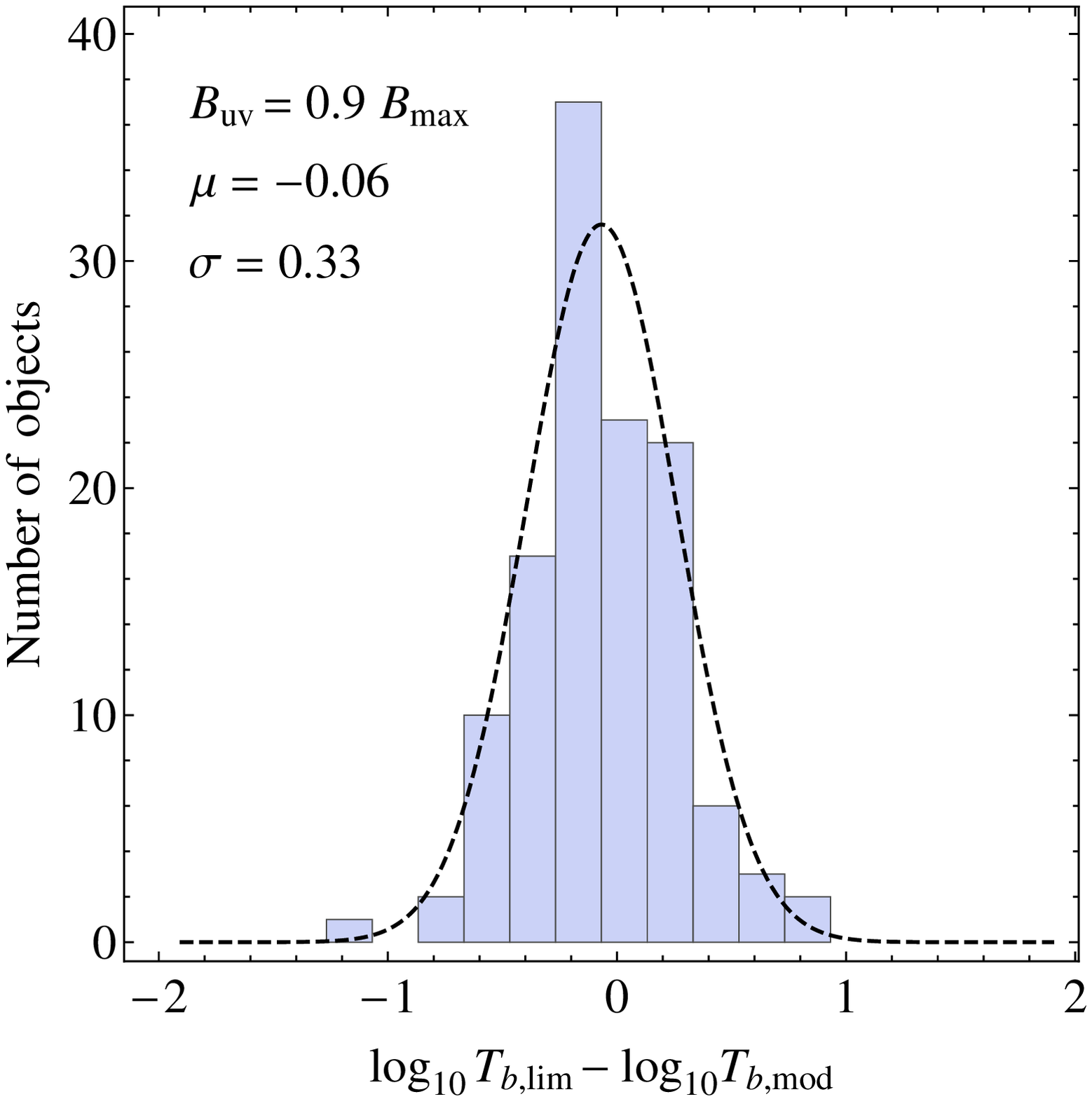}
}
\vskip 0.08\textwidth
\caption{Comparison of $T_\mathrm{b,lim}$ (circular Gaussian
  approximation) and $T_\mathrm{b,mod}$ estimates obtained from the
  86\,GHz survey data \citep{lee+2008}. Open circles indicate objects
  with lower limits on $T_\mathrm{b,mod}$. The dashed line marks the
  one-to-one correspondence between the two estimates. For each
  object, the $T_\mathrm{b,lim}$ is estimated from the 86\,GHz data at
  $R_\mathrm{uv}\ge 0.90\,B_\mathrm{max}$ to restrict the
  visibility information to the most compact structures. The resulting
  $T_\mathrm{b,lim}$ are essentially equal to $T_\mathrm{b,mod}$. The
  residual logarithmic distribution of the
  $T_\mathrm{b,lim}/T_\mathrm{b,mod}$ ratio (inset) can be approximated
  by the Gaussian PDF with $\mu =-0.06$ and $\sigma = 0.33$.}
\label{fg:86ghz}
\end{figure}

\subsection{Results from the CMVA survey data}

If $T_\mathrm{b,lim}$ is indeed determined by the limiting resolution
of the data, the difference between the $T_\mathrm{b,lim}$ and
$T_\mathrm{b,mod}$ estimates should become progressively smaller with
increasing $q_\mathrm{max} = B_\mathrm{max}/\lambda_\mathrm{obs}$ and
for model fitting performed with circular Gaussian components
(essentially forcing model fitting to be more responsive to the
resolution limit). Both these conditions are fulfilled with the
86\,GHz data from the CMVA survey.

An analysis of the 86\,GHz survey data presented in Fig.~\ref{fg:86ghz}
demonstrates that at spatial frequencies $q\ge 2$\,G$\lambda$, the
estimates $T_\mathrm{b,lim}$ and $T_\mathrm{b,mod}$ yield similar
brightness temperatures. This indicates that at these
spatial frequencies, the visibility distribution is strongly
dominated by the most compact structure, which is reflected both in
the modelfit results and in the visibility-based estimates of
$T_\mathrm{b,lim}$. It can therefore be expected that the
visibility-based estimates provide reliable brightness
temperature limits for RadioAstron space VLBI measurements
\citep{kardashev+2013} and millimeter VLBI observations
\citep{doeleman+2012}, reaching frequency spacings well in excess of
1\,G$\lambda$.

\subsection{Comparisons with $T_\mathrm{b,min}$}

Statistical comparisons of $T_\mathrm{b,mod}$ and $T_\mathrm{b,lim}$ with the
minimum brightness temperature $T_\mathrm{b,min}$ can also be made,
although the latter estimate may be affected by a general flux density
bias in the survey. We recall that $T_\mathrm{b,min}$ only depends on the
absolute value of the visibility amplitude, while $T_\mathrm{b,lim}$
effectively represents the signal-to-noise ratio (SNR) of the visibility.

The average ratios of $T_\mathrm{b,lim}/T_\mathrm{b,mod}$ are
$2.0^{+2.3}_{-1.1}$ and $1.6^{+0.7}_{-0.5}$ for the MOJAVE and CMVA
samples, respectively. These values are similar to the ratios of 2.6
and 2.1 obtained in Sect.~\ref{sc:method} from the full-track VLBI
data on 3C\,345 and NGC\,1052. The rather large spread of these
average ratios is indicative of random errors in the flux density
scales of both surveys (which can result from self-calibration applied
during imaging the survey data). The lower bounds of both averages are
$\approx 1$, which can either reflect a relatively low SNR of the
long-baseline visibilities or suggest that a systematic bias may also
affect the flux density scales. 

The average
$T_\mathrm{b,lim}/T_\mathrm{b,mod}$ ratios imply
$\mathrm{SNR}_\mathrm{MOJAVE}=3.9$ and
$\mathrm{SNR}_\mathrm{CMVA}=2.8$ on the baselines at which the
brightness temperature limits are estimated. This is indeed similar to
the measured SNR on the longest baselines in the survey data.

The average SNR can also be used to calculated the amount of bias in
the flux density scale that is needed to bring the survey results
closer to the results from the two full-track cases considered. In
both cases, a positive flux density bias of about 20\% on the longest
baselines would reconcile the survey averages with those of the
full-track data. Hence, this can also be a viable possibility, in
particular for the CMVA data that have been collected in three long
observing runs.

The comparison of $T_\mathrm{b,min}$ and $T_\mathrm{b,mod}$ obtained
from the MOJAVE data yields an average ratio
$T_\mathrm{b,min}/T_\mathrm{b,mod} = 2.3$, while a ratio of $\le 1$
is expected. This discrepancy probably results from the same
potential problems as discussed in Sect.~\ref{sc:res_mojave} in
connection with using elliptical modelfits to describe the core
region. Indeed, applying either the $\theta_\mathrm{min}$ or the
$\theta_\mathrm{lim}$ correction described in
Sect.~\ref{sc:res_mojave} reduces this ratio to 0.7. This lends
further support to the conclusion that elliptical modelfits may not be
optimal for estimating the brightness temperature in compact jets
because it is effectively determined by the transverse dimension of
the flow. Following this line of argument, the CMVA data (fitted by
circular Gaussian and providing substantially longer baselines for
estimating $T_\mathrm{b,min}$) should yield a better value of the
average $T_\mathrm{b,min}/T_\mathrm{b,mod}$ ratio. This is indeed the
case, with the average $T_\mathrm{b,min}/T_\mathrm{b,mod} = 0.5$
measured from the CMVA data.

\subsection{General implications for $T_\mathrm{b}$ measurements }

Overall, the comparisons of $T_\mathrm{b,lim}$ and $T_\mathrm{b,mod}$
made using the MOJAVE and the CMVA survey data indicate that
brightness temperature estimates obtained from the visibility flux
density at longest baselines (highest spatial frequencies) provide
suitable limits on the brightness temperature of the most compact
emitting regions in radio sources. These comparisons also suggest that
visibility-based estimates may even be more realistic in case of
complex structure of the emitting region ({\em i.e.}, a marginally
transversely resolved flow, such as a compact conical jet, or a
flow with thread-like patterns embedded).

The same situation may be realized for other types of objects studied
with interferometric measurements, for instance, in radio
interferometry observations of young supernovae \citep{marcaide+2009},
radio \citep{dyck+1998} and optical
\citep{cusano+2012,ohnaka+2013,arroyo+2014} interferometry studies of
resolved stars, and optical interferometry observations of
protoplanetary disks \citep{malbet+2005}, recurrent novae
\citep{chesneau+2007}, and active galaxies \citep{weigelt+2012}. 
  In optical interferometry experiments that measure the normalized
  visibility spectrum $V_\mathrm{q}/V_0$, the zero-spacing flux,
  $V_0$, may need to be determined in a separate measurement
  \citep[{\em cf}.][]{cusano+2012}.
In all of these cases,
brightness temperature estimates can be made with a visibility
response calculated for a specific pattern of brightness distribution
(see Appendix~A). For objects with a complex {\em internal} structure,
this would correspond to estimating a {\em pattern-averaged} brightness
temperature, while the brightness temperature of the most compact regions
can still be estimated by using the data from the longest baselines
and applying the generic Gaussian case described in Sect.~\ref{sc:method}.

The arguments presented above demonstrate that for a given
interferometric measurement, a range of plausible brightness
temperatures of the most compact emitting structure can be determined
from the visibility amplitude and its error measured at the longest
baselines contributing to the measurement. This procedure effectively 
addresses the brightness temperature of emitting regions detected at the limiting
resolution of the measurement.

In summary, our analysis leads to the
following basic conclusions about applying visibility data to
constrain the brightness temperature in astrophysical objects:
\begin{enumerate}

\item A measurement of visibility amplitude, $V_\mathrm{q}$, alone is
  sufficient to constrain the {\em minimum} brightness temperature,
  $T_\mathrm{b,min}$, that can be derived from this visibility under
  the assumption that the brightness distribution of the target object
  can be described by a Gaussian function. This holds for both circular
  and elliptical Gaussians and indeed for any visibility distribution
  that has an inflection point (for instance, for any visibility
  distribution that can be approximated by a Bessel function of the
  first kind).

\item If the brightness temperature of an emitting region with a putative Gaussian
  brightness distribution is constrained using an assumed limit on
  the zero-spacing flux density, $V_0$, the nature of the resulting
  constraint depends on the ratio $V_0/V_\mathrm{q}$. For $V_0>
  e\,V_\mathrm{q}$, lower/upper limits on $V_0$ provide lower/upper
  limits on $T_\mathrm{b}$. For $V_0< e\,V_\mathrm{q}$, lower/upper
  limits on $V_0$ provide {\em upper/lower} limits on $T_\mathrm{b}$.

\item A measurement of visibility amplitude, $V_\mathrm{q}$, and its
  r.m.s. error, $\sigma_\mathrm{q}$, can be used to calculate the {\em
    maximum} brightness temperature, $T_\mathrm{b,lim}$ that can be
  obtained under assumption that the emitting region is marginally resolved.

\item At spatial frequencies higher than $\sim 200$\,M$\lambda$, the
  interval $[T_\mathrm{b,min},T_\mathrm{b,lim}]$ provides a reliable
  bracketing for the brightness temperature of the most compact
  structure in the target object. 

\item Applying the analysis to samples of radio sources with brightness
  temperatures derived from model fitting the compact structure
  indicates that the brightness temperature in the relativistic jets
  may be largely determined by the transverse dimension of the flow. 

\item The visibility-based estimates of brightness temperature can
  offer a suitable tool for constraining this physical parameter in a
  wide range of experiments in which a full reconstruction of the source
  structure is not feasible.

\end{enumerate}

\section*{Acknowledgments}

This research has made use of data from the MOJAVE database that is
maintained by the MOJAVE team (Lister et al., 2009, AJ, 137, 3718) and
of data obtained with the Global Millimeter VLBI Array, which
consists of telescopes operated by the MPIfR, IRAM, Onsala, Metsahovi,
Yebes and the VLBA. The VLBA is an instrument of the National Radio
Astronomy Observatory, a facility of the National Science Foundation
operated under cooperative agreement by Associated Universities, Inc.

\bibliographystyle{aa}
\bibliography{tb}

\appendix

\section{Visibility limits on the brightness temperature derived for specific patterns of brightness distribution}
\label{sc:app-a}

Visibility limits on brightness temperature can be obtained by
combining the expression for brightness temperature,
\begin{eqnarray}\nonumber
T_\mathrm{b} &=& \frac{h\,\nu}{k} \ln^{-1}\left(1+ \frac{2\,h\,\nu^3}{I(r)\,c^2}\right) = \frac{h\,\nu}{k} \ln^{-1}\left(1+{\cal I}_\mathrm{b}\right)\, \\
& \rightarrow& \frac{I(r)\,c^2}{2\,k\,\nu^2}\,\quad \mathrm{for}\,\,\, h\,\nu\ll k\,T 
\label{eq:tbeq}
,\end{eqnarray}
with the visibility function, $V$, calculated for a
specific, circularly symmetric pattern of brightness distribution, $I(r)$, observed with an
interferometer at a spatial harmonic, $q$, which yields a
visibility measurement $V(q)$ and its r.m.s. error
$\sigma(q)$. Here $h$ is the Planck constant, $k$ is the
Boltzmann constant, $c$ is the light speed, and $\nu$ is the frequency
of measurement. 

For the Planck regime, the term ${\cal I}_\mathrm{r}$ is derived for
each pattern.  Full expressions for $T_\mathrm{b}$ are given for the
Rayleigh-Jeans regime.  The general form of brightness distribution
pattern is chosen to be $I(r) = \eta_\mathrm{d}\, S_\mathrm{tot}\,
f(r)$, where $S_\mathrm{tot} \equiv V(0)$ is the total or {\em
  zero-spacing} intensity of the pattern, and $f(r)$ and
$\eta_\mathrm{d}$ are the respective pattern-specific radial
dependence of brightness and its normalization factor (generally
dependent on the size $d$ of the pattern). The normalization,
$\eta_\mathrm{d}$, is chosen so that it provides a Fourier transform
${\cal F}I(r)$ of the form $V(q) = V(0)\, {\cal F}f(r)$.  To simplify
the expressions derived, $I_\mathrm{r}$, $V_0$ and ($V_\mathrm{q},
\sigma_\mathrm{q}$) denote $I(r)$, $V(0)$ and $(V(q), \sigma(q))$ in
the discussion below.

\begin{table*}[t!]
\caption{Visibility brightness temperature estimates for  specific patterns of brightness distribution}
\label{tb:app1}
\begin{center}
\small
\begin{tabular}{||l|c|c|c|c||}\hline\hline & & & \multicolumn{2}{|c||}{} \\ [-1.5ex]
 & \multicolumn{1}{c|}{Circular Gaussian} & \multicolumn{1}{c|}{Uniformly bright} & \multicolumn{2}{c||}{Optically thin shell of finite thickness} \\ [1.0ex] \cline{4-5}& & & & \\ [-1.5ex]
 & \multicolumn{1}{c|}{component} & \multicolumn{1}{c|}{disk} & \multicolumn{1}{c|}{shell average} & \multicolumn{1}{c||}{peak brightness} 
\\[0.5ex] \hline\hline & & & \multicolumn{2}{|c||}{} \\ [-1.5ex]
$I_\mathrm{r}$ 
& \(\ds \frac{2\,\sqrt{\ln 2}} {\sqrt{\pi}}\,\frac{S_\mathrm{tot}}{d} \exp\left(-\frac{4\,\ln 2 r^2}{d^2}\right)\) 
& \(\ds S_\mathrm{tot}
\begin{cases}
4/(\pi\,d^2), & r\le d/2\,,\\
0 & r> d/2
\end{cases} \)   
& \multicolumn{2}{l||}
{\(\ds I_0
\begin{cases}
 (d^2-4r^2)^{1/2}-(\alpha^2 d^2 - 4r^2)^{1/2} & 2r< \alpha\,d\,,\\
 (d^2-4r^2)^{1/2} & \alpha\,d \le 2r \le d\,,\\
0 & 2r>d\,,\\
\end{cases} \)
}
\\
 & & & \multicolumn{2}{c||}{with \(\ds I_0 =\frac{6 S_\mathrm{tot}}{\pi\,d^3 (1-\alpha^3)} \) }
\\[2.5ex]\hline & & & & \\ [-1.5ex]
%
$T_\mathrm{b}$ 
& \(\ds \frac{2\,\ln\,2}{\pi\,k}\,\frac{S_\mathrm{tot}\,\lambda^2}{d^2} \)
& \(\ds \frac{2}{\pi\,k}\frac{S_\mathrm{tot}\,\lambda^2}{d^2}  \)
& \(\ds \frac{2}{\pi\,k}\,\frac{S_\mathrm{tot}\,\lambda^2}{d^2}  \)
& \(\ds \frac{3}{\pi\,k}\,\frac{S_\mathrm{tot}\,\lambda^2}{d^2\,(1-\alpha^3)^{1/2}}  \)
\\[2.5ex]\hline & & & & \\ [-1.5ex]
%
${\cal I}_\mathrm{b}$ 
& \(\ds \frac{\pi\,h\,\nu}{2\,\ln\,2}\,\frac{d^2}{S_\mathrm{tot}\,\lambda^2} \)
& \(\ds \frac{\pi\,h\,\nu}{2}\,\frac{d^2}{S_\mathrm{tot}\,\lambda^2} \)
& \(\ds \frac{\pi\,h\,\nu}{2}\,\frac{d^2}{S_\mathrm{tot}\,\lambda^2}  \)
& \(\ds \frac{\pi\,h\,\nu}{3}\,\frac{d^2\,(1-\alpha^3)^{1/2}}{S_\mathrm{tot}\,\lambda^2}  \)
\\[2.5ex]\hline & & & \multicolumn{2}{|c||}{} \\ [-1.5ex]
%
$V_\mathrm{q}$ 
& \(\ds V_0\, \exp\left(-\frac{\pi^2\,q^2\,d^2}{4\,\ln 2}\right)   \)
& \(\ds V_0 \, \frac{2\,J_{1}(\pi\,d\,q)}{\pi\,d\,q} \approx \)
& \multicolumn{2}{l||}{
 \(\ds V_0\, \frac{3}{(1-\alpha^3)\pi^3 d^3 q^3}\left[ \sin(\pi\,d\,q) - \right.
\pi\,d\,q \cos(\pi\,d\,q) -\)}
\\[2.5ex]
 & & & \multicolumn{2}{r||}{\(\ds \left. - \sin(\alpha\,\pi\,d\,q) + \alpha\,\pi\,d\,q \cos(\alpha\,\pi\,d\,q) \right] \approx \)  }
\\[2.5ex]
 & & \multicolumn{1}{c|}{\(\ds \approx V_0 \left(1 - \frac{\pi^2\,d^2\,q^2}{8}\right)  \)} & \multicolumn{2}{c||}{\(\ds \approx V_0 \left(1- \frac{\pi^2 d^2 q^2}{10} \frac{1 - \alpha^5}{1-\alpha^3}
\right)\)
}
\\[2.5ex]\hline & & & \multicolumn{2}{|c||}{} \\ [-1.5ex]
%
$d_\mathrm{q}$ 
& \(\ds \frac{2 \sqrt{\ln 2}}{\pi}\frac{1}{q}\sqrt{\ln(V_0/V_\mathrm{q})}  \)
& \(\ds \frac{2\,\sqrt{2}}{\pi\,q}\,\sqrt{1-V_\mathrm{q}/V_0}  \)   
& \multicolumn{2}{c||}{\(\ds \frac{\sqrt{10}}{\pi\,q} \, \left(\frac{1 - \alpha^3}{1-\alpha^5}\right)^{1/2}
\sqrt{1-V_\mathrm{q}/V_0}  \)}
\\[2.5ex]\hline & & & & \\ [-1.5ex]
%
$T_\mathrm{b,q}$ 
& \(\ds \frac{\pi}{2 k} \frac{B^2\,V_0}{\ln(V_0/V_\mathrm{q})} \)
& \(\ds \frac{\pi}{4\,k}\frac{B^2\, V_0^2}{V_0-V_\mathrm{q}}   \)   
& \(\ds \frac{\pi}{5\,k}\, \frac{1 - \alpha^5}{1-\alpha^3}\, \frac{B^2\,V_0^2}{V_0-V_\mathrm{q}}  \)
& \(\ds \frac{3\pi}{10\,k}\, \frac{1 - \alpha^5}{(1-\alpha^3)^{3/2}}\, \frac{B^2\,V_0^2}{V_0-V_\mathrm{q}}  \)
\\[2.5ex]\hline & & & & \\ [-1.5ex]
%
${\cal I}_\mathrm{b,q}$ 
& \(\ds \frac{2\,h\,\nu}{\pi}\,\frac{\ln(V_0/V_\mathrm{q})}{B^2\, V_0}  \)
& \(\ds \frac{4\,h\,\nu}{\pi}\,\frac{V_0 - V_\mathrm{q}}{B^2 \,V_0^2}  \)   
& \(\ds \frac{5\,h\,\nu}{\pi}\,\frac{1 - \alpha^3}{1-\alpha^5}\,\frac{V_0 - V_\mathrm{q}}{B^2 \,V_0^2} \)
& \(\ds \frac{10\,h\,\nu}{3\,\pi}\,\frac{(1 - \alpha^3)^{3/2}}{1-\alpha^5}\, \,\frac{V_0 - V_\mathrm{q}}{B^2 \,V_0^2}  \)
\\[2.5ex]\hline & & & & \\ [-1.5ex]
%
$T_\mathrm{b,min}$ 
& \(\ds \frac{\pi\,e}{2 k}\, B^2\, V_\mathrm{q}  \)
& \(\ds \frac{\pi}{k}\,B^2\, V_\mathrm{q}  \)   
& \(\ds \frac{4 \pi}{5\,k}\, \frac{1 - \alpha^5}{1-\alpha^3}\, B^2\,V_\mathrm{q}  \)
& \(\ds \frac{6 \pi}{5\,k}\, \frac{1 - \alpha^5}{(1-\alpha^3)^{1/2}}\, B^2\,V_\mathrm{q}  \)
\\[2.5ex]\hline & & & & \\ [-1.5ex]
%
${\cal I}_\mathrm{b,min}$ 
& \(\ds \frac{2\, h\,\nu}{\pi\, e\, B^2\, V_\mathrm{q}}  \)
& \(\ds \frac{h\,\nu}{\pi\,B^2\,V_\mathrm{q}}  \)   
& \(\ds \frac{5\,h\,\nu}{4\,\pi}\,\frac{1 - \alpha^3}{1-\alpha^5}\,\frac{1}{B^2 \,V_\mathrm{q}}  \)
& \(\ds \frac{5\,h\,\nu}{6\,\pi}\,\frac{(1 - \alpha^3)^{3/2}}{1-\alpha^5}\,\frac{1}{B^2 \,V_\mathrm{q}} \)
\\[2.5ex]\hline & & & & \\ [-1.5ex]
%
$T_\mathrm{b,lim}$ 
& \(\ds \frac{\pi}{2\,k} \, \frac{B^2\,(V_\mathrm{q}+\sigma_\mathrm{q})}{\ln [(V_\mathrm{q} + \sigma_\mathrm{q})/V_\mathrm{q}]}  \)
& \(\ds \frac{\pi}{4\,k}\, \frac{B^2\,(V_\mathrm{q}+\sigma_\mathrm{q})^2}{\sigma_\mathrm{q}}  \)   
& \(\ds \frac{\pi}{5\,k}\, \frac{1 - \alpha^5}{1-\alpha^3}\, \frac{B^2\,(V_\mathrm{q}+\sigma_\mathrm{q})^2}{\sigma_\mathrm{q}}  \)
& \(\ds \frac{3\pi}{10\,k}\frac{1 - \alpha^5}{(1-\alpha^3)^{3/2}}\frac{B^2 (V_\mathrm{q}+\sigma_\mathrm{q})^2}{\sigma_\mathrm{q}}  \)
\\[2.5ex]\hline & & & & \\ [-1.5ex]
%
${\cal I}_\mathrm{b,lim}$ 
& \(\ds \frac{2\,h\,\nu}{\pi}\,\frac{\ln[(V_\mathrm{q}+\sigma_\mathrm{q})/V_\mathrm{q}]}{B^2\, (V_\mathrm{q}+\sigma_\mathrm{q})}  \)
& \(\ds \frac{4\,h\,\nu}{\pi}\,\frac{\sigma_\mathrm{q}}{B^2\,(V_\mathrm{q}+\sigma_\mathrm{q})^2}  \)   
& \(\ds \frac{5\, h\,\nu}{\pi}\,\frac{1 - \alpha^3}{1-\alpha^5}\,\frac{\sigma_\mathrm{q}}{B^2\,(V_\mathrm{q}+\sigma_\mathrm{q})^2}  \)
& \(\ds \frac{10\,h\,\nu}{3\,\pi}\,\frac{(1 - \alpha^3)^{3/2}}{1-\alpha^5}\, \,\frac{\sigma_\mathrm{q}}{B^2\,(V_\mathrm{q}+\sigma_\mathrm{q})^2}  \)
\\[2.5ex]\hline\hline
\end{tabular}
\end{center}
\normalsize
{\bf Notes:} $I_\mathrm{r}$ -- brightness distribution function; 
$T_\mathrm{b}$ -- brightness temperature estimated from $I_\mathrm{r}$; 
$V_\mathrm{q}$ -- visibility function; $d_\mathrm{q}$ -- pattern size, estimated 
from the visibility ratio $V_\mathrm{q}/V_0$; $T_\mathrm{b,q}$ -- brightness 
temperature corresponding to $d_\mathrm{q}$; $T_\mathrm{b,min}$ -- minimum 
brightness temperature, obtained with $V_0 = e\,V_\mathrm{q}$ for the Gaussian 
component and $V_0 = 2\,V_\mathrm{q}$ for the disk and the shell; 
$T_\mathrm{b,lim}$ -- limiting brightness temperature, obtained with $V_0 = V_\mathrm{q}+\sigma_\mathrm{q}$. For the brightness temperature estimates, the respective ${\cal I}$ expression give the
${\cal I}_\mathrm{b}$ term in the Planck form of the brightness temperature equation.
\end{table*}

With the adopted normalization, a generic solution for the size
$d_\mathrm{q}$ of the pattern can be determined by solving ${\cal
  F}\,I_\mathrm{r} = V_\mathrm{q}/(V_0)$ for $d$. The minimum
brightness temperature supported by the visibility measurement
$V_\mathrm{q}$ is obtained by substituting $I_\mathrm{r}$ with
$I_\mathrm{r}(d$$\rightarrow$$d_\mathrm{q})$ in Eq.~\ref{eq:tbeq}
and finding the value of $V_0$ that minimizes the respective
$T_\mathrm{b}$ given by this equation ({\em e.g.}, deriving $V_0$ from the 
condition $\mathrm{d} T_\mathrm{b}/\mathrm{d} V_0 = 0$).

To derive the maximum measurable brightness temperature, the
characteristic size, $d_\mathrm{lim}$, of the pattern is obtained from the
relation ${\cal F}I_\mathrm{r} =
V_\mathrm{q}/(V_\mathrm{q}+\sigma_\mathrm{q})$. The maximum measurable
brightness temperature $T_\mathrm{b,lim}$ is then calculated by
substituting $I_\mathrm{r}$ in Eq.~\ref{eq:tbeq} with
$I_\mathrm{r}(d$$\rightarrow$$d_\mathrm{lim})$. 

Results of the calculations are presented in Table~\ref{tb:app1} for a
circular Gaussian component, a disk of uniform brightness, and a
spherical shell of finite thickness. The characteristic size $d$
represents, respectively, the full width at half maximum of the
Gaussian component, the diameter of the disk, and the outer diameter
of the shell. The spherical shell is further defined by its thickness
$\delta_\mathrm{r} =(1-\alpha) d/2$ (with $0\le\alpha\le 1$).  Thus,
$\alpha=0$ describes an infinitely thin shell and $\alpha\rightarrow
1$ describes a filled, optically thin sphere.  Calculations for the
spherical shell are made separately for the average shell brightness and for the
peak brightness $I_\mathrm{r,peak}$ in the shell, realized at
$r=\alpha\,d/2$, with the resulting $I_\mathrm{r,peak} =
6\,S_\mathrm{tot}/\pi\,d^2 \,(1-\alpha^3)^{1/2}$.

\end{document}